\documentclass[twocolumn, prb,superscriptaddress]{revtex4-1}
\usepackage{amsmath}
\usepackage{float}
\usepackage{epstopdf}
\usepackage{verbatim}
\usepackage{amsfonts,amssymb}
\usepackage{graphicx}
\usepackage{hyperref}    % links
\usepackage{bm}          % bold math letters
\usepackage{natbib}
\usepackage{soul} % in order to highlight
\usepackage{color}
\usepackage{longtable}
\usepackage{ textcomp }

\begin{document}

\title{Electron gas induced in SrTiO$_3$}

\author{Han Fu}
\email{fuxxx254@umn.edu}
\affiliation{Fine Theoretical Physics Institute, University of Minnesota, Minneapolis, MN 55455, USA}
\author{K. V. Reich}
\affiliation{Fine Theoretical Physics Institute, University of Minnesota, Minneapolis, MN 55455, USA}
\affiliation{Ioffe Institute, St Petersburg, 194021, Russia}
\author{B. I. Shklovskii}
\affiliation{Fine Theoretical Physics Institute, University of Minnesota, Minneapolis, MN 55455, USA}

\begin{abstract}
This mini-review is dedicated to the 85th birthday of Prof. L. V. Keldysh, from whom we have learned so much. In this paper we study the potential and electron density depth profiles in surface accumulation layers in crystals with a large and nonlinear dielectric response such as SrTiO$_3$ (STO) in the cases of planar, spherical and cylindrical geometries. The electron gas can be created by applying an induction $D_0$ to the STO surface. We describe the lattice dielectric response of STO using the Landau-Ginzburg free energy expansion and employ the Thomas-Fermi (TF) approximation for the electron gas. For the planar geometry we arrive at the electron density profile $n(x) \propto (x+d)^{-12/7}$, where $d \propto D_0^{-7/5} $. We extend our results to overlapping electron gases in GTO/STO/GTO heterojunctions and electron gases created by spill-out from NSTO (heavily $n$-type doped STO) layers into STO. Generalization of our approach to a spherical donor cluster creating a big TF atom with electrons in STO brings us to the problem of supercharged nuclei. It is known that for an atom with nuclear charge $Ze$, where $Z > 170$, electrons collapse onto the nucleus resulting in a net charge $Z_n < Z$. Here, instead of relativistic physics, the collapse is caused by the nonlinear dielectric response. Electrons collapse into the charged spherical donor cluster with radius $R$ when its total charge number $Z$ exceeds the critical value $Z_c \simeq R/a$, where $a$ is the lattice constant. The net charge $eZ_n$ grows with $Z$ until $Z$ exceeds $Z^* \simeq (R/a)^{9/7}$. After this point, the charge number of the compact core $Z_n$ remains $\simeq Z^*$, with the rest $Z^*$ electrons forming a sparse Thomas-Fermi atom with it. We extend our studies of collapse to the case of long cylindrical clusters as well.
\end{abstract}

\date{\today}

\maketitle

\section{Introduction}

In recent years, there has been growing interest in the investigation of $\mathrm{ABO_3}$ perovskite crystals, which are important for numerous technological applications and show intriguing magnetic, superconducting, and multiferroic  properties \cite{Oxides_rev}.  Special attention \cite{Stemmer_STO,Zubko_oxides} is paid to heterostructures involving $\mathrm{SrTiO_3}$ (STO) which is a  semiconductor  with a band gap of $\simeq \mathrm{3.2~eV}$  \cite{Optical_absorbtion_STO} and a large dielectric constant $\kappa$ ranging from $2 \cdot 10^4$ at liquid helium temperatures to $350$ at room temperature.  As with conventional semiconductors,  $\mathrm{SrTiO_3}$ can be used as a building block for different types of devices, with reasonably large mobility \cite{Ohtomo_2004,Hwang_mobility}.

\begin{figure}
\includegraphics[width=0.8\linewidth]{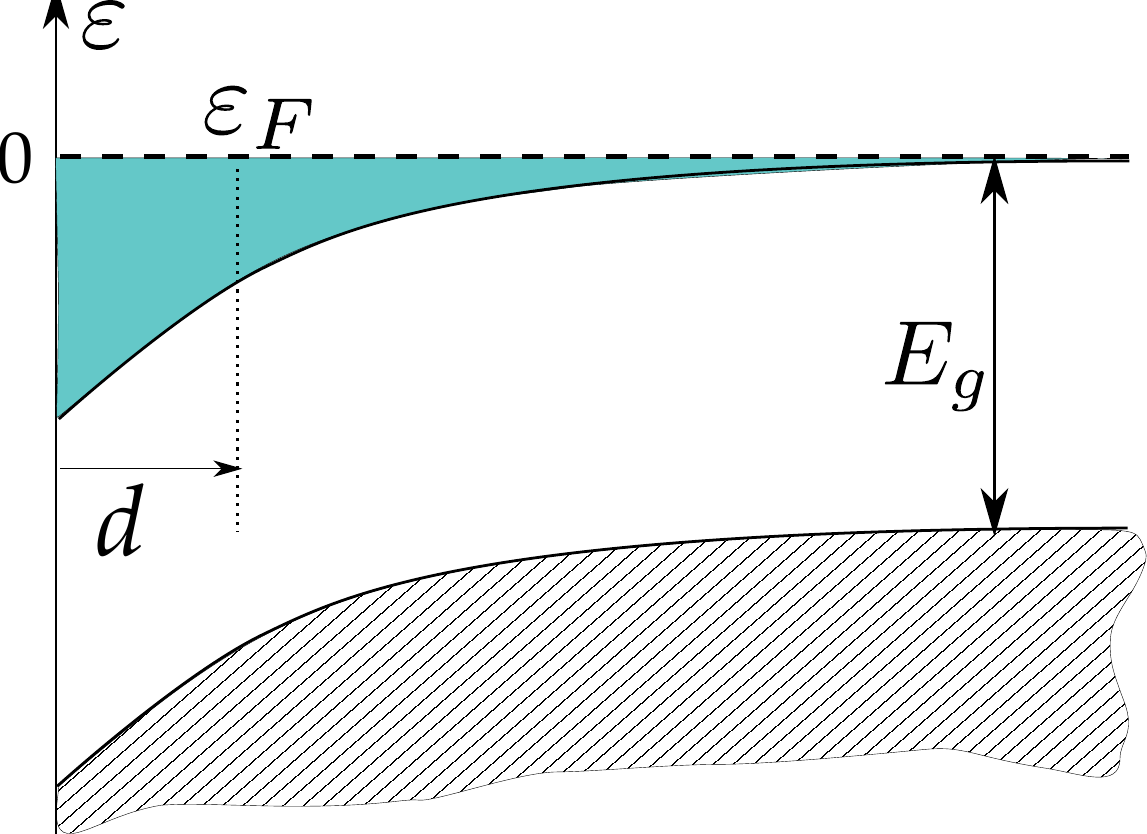}
\caption{(Color  online) Schematic energy diagram of an accumulation layer in a moderately $n$-doped STO. The band gap is $E_g$. Electrons (blue region) are attracted by an external induction $D_0$. The characteristic width of the electron gas is $d$. In the bulk of STO the Fermi level   $\varepsilon_F $ is near the bottom of the conduction band  (plotted by the dashed line)}
\label{fig:accumulation}
\end{figure}

Many devices are based on the accumulation layer of electrons  near a heterojunction interface involving a moderately $n$-type doped STO. For example, one can get an electron accumulation layer on the STO side of the GTO/STO heterojunction induced by the electric field resulting from the ``polar catastrophe" in GdTiO$_3$ (GTO) \cite{polar-catastrophe_2006} (see Fig. \ref{fig:accumulation}). The role of GTO can also be played by perovskites LaAlO$_3$ \cite{Ohtomo_2004,Hwang_mobility,Stemmer_STO}, NdAlO$_3$, LaVO$_3$ \cite{LaVO_STO}, PrAlO$_3$, NdGaO$_3$ \cite{different_polar_STO}, LaGaO$_3$ \cite{LaGaO_STO}, LaTiO$_3$ \cite{LaTO_STO} and others producing the polar catastrophe \cite{polar-catastrophe_2006}. One can  accumulate an electron gas using a field effect \cite{10_percent,Hwang_gate,Stemmer_concentration_interface}. In Refs. \onlinecite{induced_superconductivity,Gallagher_2014} the authors accumulated up to $10^{14} ~\mathrm{cm}^{-2}$ electrons on the surface of STO  using ionic liquid gating. Inside bulk STO $\delta$-doping can be used  to introduce two accumulation layers of electrons  \cite{delta_doped_stemmer,delta_doped_STO_Hwang,delta_doped_STO_Stemmer}. It is natural to think that the depth profiles of the potential and electron density inside STO have a universal origin in all these devices.

Interface properties determine characteristics of  all these devices. Not surprisingly, the potential and electron density depth profiles in such devices have attracted attention from the experimental  \cite{Hwang_Xray,Hwang_PL,LAO_STO_Berreman,Stemmer_GdTO} and theoretical points of view \cite{MacDonald_theory,induced_superconductivity, abinitio_STO,abinitio_STO_2,distribution_LAO_STO,superconductivity_LAO_STO}. For example, experimental data show that electrons are distributed in a layer of width $\simeq \mathrm{5-10~nm}$  near the   $\mathrm{LaAlO_3 / SrTiO_3}$ interface. Theoretical works that attempt to explain such behavior are based on microscopic numerical calculations.

The goal of this minireview based on Refs. \onlinecite{HRS, RS} is to create a simple, mostly phenomenological approach for describing the electron density depth profiles in STO. To account for the nonlinear dielectric response in STO we use the Landau-Ginzburg free energy expansion \cite{Ginzburg_ferroelectrics, Landau_stat}. Electrons are almost everywhere described in the Thomas-Fermi approximation \cite{Thomas_Fermi}.  Although we mostly concentrate on accumulation layers in STO, the developed approach is applicable to $\mathrm{KTaO_3}$ \cite{Rowley_2014} and $\mathrm{CaTiO_3}$ \cite{CaTiO3_quantum} serving as the host media for accumulation layers as well.

Our main result is a new form for the potential and electron density depth profiles in various planar STO structures due to the nonlinear dielectric response. In particular, for an accumulation layer in STO created by an external interfacial induction $D_0$, we find an electron concentration $n(x)$ that depends on the distance from the surface $x$ as $n(x) \propto (x+d)^{-12/7}$, where the width $d$ decreases with $D_0$ as $d \propto D_0^{-7/5}$. It is shown in Ref. \onlinecite{RS} that these relations seem to agree with experimental data \cite{LAO_STO_Berreman,Hwang_PL}. Using this basic solution for a separate accumulation layer we study more complicated problems where accumulation layers overlap, e.g., the structure of GTO/STO/GTO multi-heterojunctions \cite{Stemmer_2012, Stemmer_2013,Stemmer_2015}. We calculate how electron density profiles evolve as a function of the the distance between two heterojunctions. Another planar problem arises when the accumulation layer is created in STO which is a part of the modulation doped structure of NSTO/STO \cite{Hwang_2009,Stemmer_2010,Ohtomo_2002,Chang_2013,Choi_2012}. In this case electrons spill out from the heavily $n$-type doped STO layer (NSTO) and the induction $D_0$ is self-consistently  provided by the depletion layer in NSTO.

The experience with above planar problems allows us to study the electron gas in STO created by the external charges or electric field with spherical and cylindrical symmetry. Such devices cab be realized by doping the bulk STO by generating oxygen vacancies at high temperatures. The vacancies either form along a network of extended defects\cite{Ov2002} or assemble together to lower the system's energy\cite{Ov2007, Muller_2004}, producing large positively charged donor clusters.

Another way to more controllably create such a cluster is to ``draw" a disc of charge by the atomic force microscope (AFM) tip on the surface of LAO/STO structure with the subcritical thickness for LaAlO$_3$ (LAO) \cite{Cen_2008,Cen_2010}. The potential caused by such a positive disc in the bulk STO is similar to that of a charged sphere.

Let us consider a spherical donor cluster with radius $R$ and charge $Ze$. At relatively small $Z$, there are $Z$ electrons located at distances from $\kappa b/Z$ to the Bohr radius $\kappa b$ from the cluster, which form a Thomas-Fermi ``atom" \cite{Landau} with it. Here $b=\hbar^2/m^*e^2$, $m^*\approx1.5m$ is the effective electron mass in STO \cite{RS} with $m$ being the free electron mass. Since $\kappa$ is large, the electrons are far away from the cluster and the whole ``atom" is very big. As $Z$ increases, the electron gas swells inward to hold more electrons. However, we find that when $Z$ goes beyond a certain value $Z_c$ ($\kappa b/Z$ is still much larger than $R$ at this moment), the physical picture is qualitatively altered. Surrounding electrons start to collapse into the cluster and the net cluster charge gets renormalized from $Ze$ to $Z_ne$ with $Z_n\ll Z$ at very large $Z$.

The effect of charge renormalization is not new \cite{Pomeranchuk, *Zeldovich, Kolomeisky}. For a highly charged atomic nucleus with charge $Ze$, the vacuum is predicted to be unstable against creation of electron-positron pairs, resulting in a collapse of electrons onto the nucleus with positrons emitted \cite{Pomeranchuk,Zeldovich}. This instability happens when $Z>Z_c$ with $Z_c\simeq170\gtrsim1/\alpha$, where $\alpha=e^2/\hbar c\simeq1/137$ is the fine structure constant. When $Z$ exceeds $Z^*\simeq1/\alpha^{3/2}\simeq137^{3/2}$, the net charge saturates at $Z^*$ (see Ref. \onlinecite{Kolomeisky}). In the condensed matter setting, there are similar phenomena in narrow-band gap semiconductors and Weyl semimetals \cite{Kolomeisky} as well as graphene \cite{Fogler, *Novikov, *Levitov,*Pereira, *Gorsky, * Crommie}.  In all these cases, the collapse happens because the energy dispersion of electrons is relativistic in the strong Coulomb field of a compact donor cluster playing the role of a nucleus. In our work, however, the collapse originates from the strong nonlinearity of dielectric constant in STO at small distances from the cluster. In the case of a spherical donor cluster, this nonlinearity leads to the change of the attractive potential near the cluster from being $\propto 1/r$ to $\propto1/r^5$, resulting in the collapse of non-relativistic electrons to the cluster.

\begin{figure}[h]
$\begin{array}{c}
\includegraphics[width=0.47\textwidth]{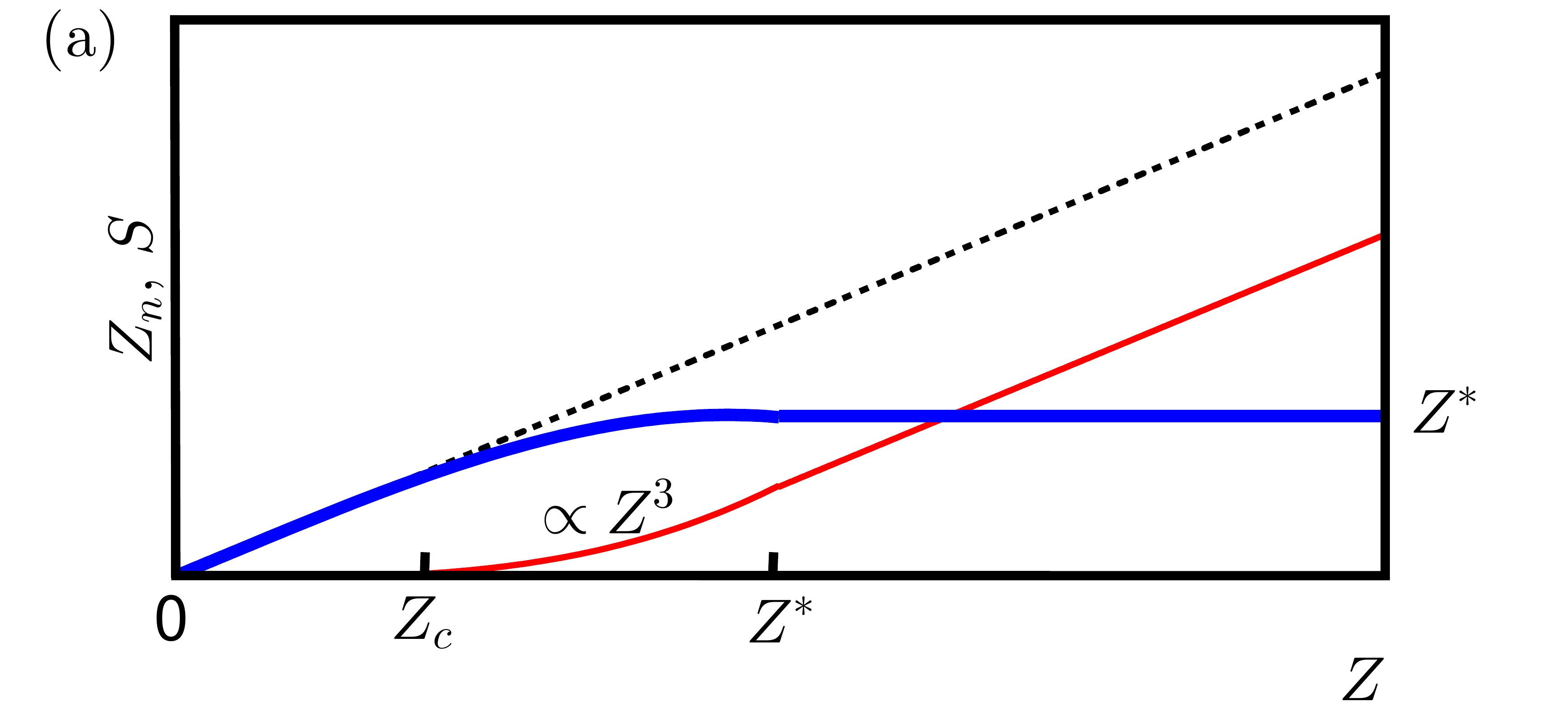}\\
\\
\\
\includegraphics[width=0.47\textwidth]{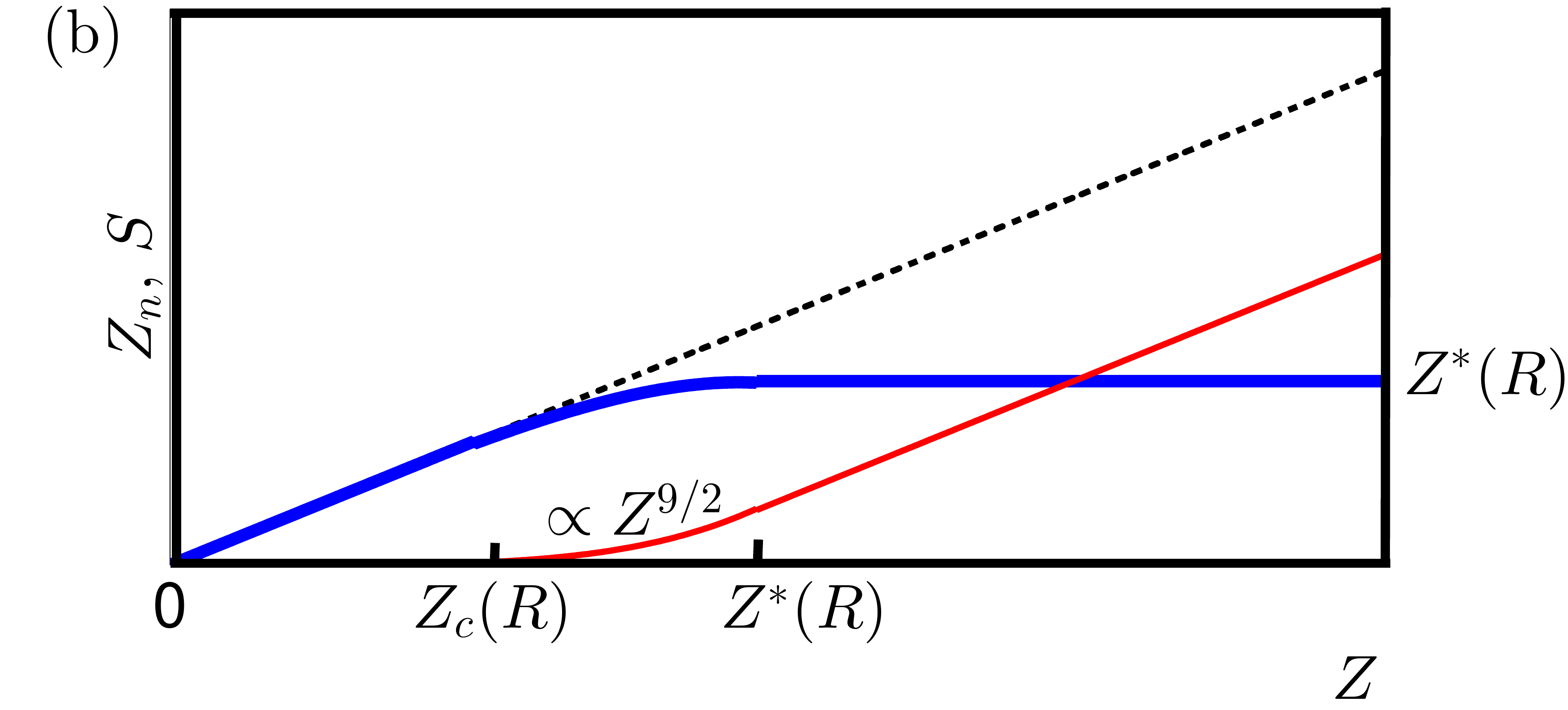}
\end{array}$
\caption{(Color online) The number of collapsed electrons $S$ and the renormalized net charge $Z_ne$ as a function of the original charge $Ze$. $S$ is shown by the thin solid line (red), $Z_n$ is denoted by the thick solid line (blue), and the dashed line (black) is a guide-to-eye where $Z_n=Z$. $Z_c$ denotes the critical value where electrons begin to collapse and $Z^*$ is the saturation point where $Z_n$ stops growing. (a) Collapse of electrons and charge renormalization for highly charged nuclei. $S\propto Z^3$ at $Z_c\ll Z\ll Z^*$ (see Ref. \onlinecite{Migdal_1977}). (b) Collapse of electrons and charge renormalization for spherical donor clusters in STO. $S\propto Z^{9/2}$ at $Z_c\ll Z\ll Z^*$.}\label{fig:1}
\end{figure}
The phenomena of electron collapse and charge renormalization in both heavy nuclei and our work are presented in Fig. \ref{fig:1}. In our case, the first electron collapses at $Z\simeq Z_c\simeq R/a$, where $a$ is the lattice constant, and at $Z\gg Z^*\simeq (R/a)^{9/7}$, the net charge of nucleus $Z_ne$ saturates as $Z_n\simeq Z^*$.

We use the Thomas-Fermi approximation to show how the electron gas collapses into the cluster at $Z\gg Z_c$ and find the corresponding electron density profile of the new two-scale Thomas-Fermi atom (see Fig. \ref{fig:3a}).

The remainder of this paper is organized as follows. In Sec.~\ref{sec:model} we define the model based on the Landau-Ginzburg theory for calculating the lattice dielectric response and describe the parameters STO. In Sec.~\ref{sec:singlep} we use the Thomas-Fermi approach to calculate the electron accumulation layer for a single GTO/STO heterojunction induced by the polar catastrophe. In the Sec.~\ref{sec:overlap} we consider the case of multiple heterojunctions where accumulation layers from different interfaces overlap. In Sec. \ref{sec:spillout} we consider the case in which the electrons spill out from a heavily $n$-type doped STO material (NSTO) to the moderately $n$-type doped one for both a single junction and multi-junctions. In Secs.~\ref{sec:spherical} and \ref{sec:cylindrical} we extend our studies to the non-planar structure consisting of spherical or cylindrical donor clusters inside STO. In Sec. \ref{sec:temp}, we go beyond the zero temperature and investigate the thermal ionization of spherical donor clusters and its experimental implications. Sec. ~\ref{sec:conclusion} provides a summary and conclusion.

\section{The model}
\label{sec:model}

Bulk STO typically is an $n$-type semiconductor with a concentration of donors  $N>10^{17} ~ \mathrm{cm^{-3}}$. Let us discuss the position of Fermi energy $\varepsilon_F$ in such crystals. The electron spectrum near the bottom of the conduction band is complicated~\cite{Mazin_band_structure}, and in order to make the problem of an accumulation layer tractable analytically we assume that it is isotropic and non-degenerate with the effective mass $m^* \simeq 1.5~ m$, \cite{RS} where $m$ is free electron mass. Within the hydrogenic theory of
shallow donors, the donor Bohr radius is equal to  $\kappa b$, where $ b = \hbar^2/m^* e^2 \simeq 0.35 ~\mathrm{\AA}$, $e$ is the electron charge, and $\kappa$ is dielectric constant of the material. At room temperature when $\kappa = 350$, the Bohr radius  $\kappa b = \mathrm{123 ~ \AA}$ is so large that the Mott criterion for the metal-insulator transition in doped semiconductors $N_c b^3 =0.02/\kappa^3$ leads to a very small critical concentration $N_c = 1\cdot 10^{16} ~\mathrm{cm^{-3}}$. At helium temperatures $\kappa=2 \cdot 10^{4}$ and $N_c =6 \cdot 10^{10} ~ \mathrm{cm^{-3}} $. Thus, at the experimentally relevant
concentration of donors  $N>10^{17} ~\mathrm{cm^{-3}}$, we are dealing with a moderately doped semiconductor in which the Fermi energy lies in the conduction band of STO. On the other hand, due to the relatively high effective mass the bulk Fermi energy $\varepsilon_F$ is smaller than the bending energy of the conduction band bottom near the interface (see Fig. \ref{fig:accumulation}). For example, for $N =10^{18} \mathrm{cm^{-3}}$, the low temperature Fermi energy calculated from the bottom of the conduction band is $\varepsilon_F \simeq 4~\mathrm{meV},$ which can be up to $100$ times smaller than the bending energy of the conduction band bottom in an accumulation layer for GTO/STO. Therefore, we assume below that the Fermi energy coincides with the bottom of the conduction band.

We are interested in the electron distribution near an interface of STO. We consider the case when the axis $x$ is directed perpendicular to the interface  (plane $x=0$) and lies along the [100] axis of a cubic crystal of STO. (In fact, STO changes symmetry  from cubic to tetragonal at  $T \simeq 110 K$, but the distortion is small \cite{STO_lattice} and can be neglected). An external induction $D_0$ applied from the left (see Fig. \ref{fig:accumulation}) is directed along the $x$ axis. In that case the problem is effectively one-dimensional. If the charge density  is denoted by $\rho(x)$, then  the potential depth profile $\varphi(x)$ in the system is determined by the  equations:

\begin{eqnarray}
  \label{eq:Gauss}
  \frac{d D}{dx}   = 4 \pi  \rho ,  ~  D =E+4\pi P , ~   \frac {d \varphi}{d x}  = -E ,
\end{eqnarray}
where $D(x),\, E(x),\, P(x)$ are the electric displacement field (the induction), the electric field and the electric polarization in STO.  Equations (\ref{eq:Gauss}) should be solved with proper boundary conditions. For example, for a single accumulation layer the boundary conditions are $D(0)=D_0$ and $\varphi(\infty)=0$.

To solve the system (\ref{eq:Gauss})  one needs to know two material relationships  $E(P)$ and $\rho(\varphi)$. Let us start from the lattice dielectric response $E(P)$. STO is well known as a quantum paraelectric, where the onset of ferroelectric order is suppressed by quantum fluctuations \cite{Ferroelectricity_by18O}.

A powerful approach to describe the properties of ferroelectric-like materials is based on the Landau-Ginzburg theory. For a continuous second-order phase transition the Landau-Ginzburg expression of the free energy density $F$ is represented as a power series expansion with respect to the polarization $P$:
\begin{equation}
  \label{eq:F}
  F= F_0 +\frac{\tau}{2} P^2 + \frac{1}{4} A \frac{1}{P_0^2} P^4  -E P,
\end{equation}
where $F_0$ stands for the free energy density at $P = 0$ and $\tau$ is the inverse susceptibility $\tau=4\pi/(\kappa-1) \simeq 4\pi/\kappa$. In this work $0<\tau \ll 1$, $P_0=e/a^2$ is the characteristic polarization and $a \simeq \mathrm{3.9 ~\AA}$ \cite{STO_lattice} is the lattice constant. The coefficient A  describes the non-linear dielectric response. For all estimates below we use $A=0.8$ following from Ref. \onlinecite{RS}. The last term of Eq. (\ref{eq:F}) is responsible for the interaction between the polarization and the electric field $E$. In general $F$  depends on the components of the vector $P$, but in the chosen geometry  the problem is one-dimensional, and all vectors are directed along the $x$ axis.
The crystal polarization $P$ is determined by minimizing the free energy density $F$ in the presence of the electric field $E$, $\delta F/ \delta P=0$. This condition  relates $E$ and $P$,

\begin{equation}
  \label{eq:electric_field_definition}
 E = \frac{4\pi}{\kappa} P + \frac{A}{P_0^2} P^3.
\end{equation}
We note that $E \ll 4\pi P$ and thus $D=E+4\pi P \simeq 4\pi P$. The induction $D_c$ at which the transition from linear to nonlinear dielectric response  occurs can be found by equating the first and second terms in the expression (\ref{eq:electric_field_definition}):
\begin{equation}
  \label{eq:Dc}
D_c = P_0 \sqrt{\frac{(4\pi)^3}{\kappa A}}.
\end{equation}

If $D \ll D_c $  the dielectric response of STO is linear and one can use the simplified expression for the electric field:
\begin{equation}
  \label{eq:electric_field_small}
 E =  \frac{D}{\kappa}.
\end{equation}

For $D \gg D_c $  the dielectric response of STO is nonlinear and one must instead use the expression:

\begin{equation}
  \label{eq:electric_field_high}
E= \frac{A}{(4\pi)^3 P_0^2} D^3.
\end{equation}
Next one should specify $\rho(\varphi)$, which depends on the specific device of interest.

\section{A single accumulation layer}
\label{sec:singlep}
In a single heterojunction the external induction $D_0$ caused by the ``polar catastrophe" on the interface attracts electrons with a three-dimensional concentration $n(x)$ inside the accumulation layer of STO (see Fig. \ref{fig:accumulation}). Our goal is to find the electron depth profile $n(x)$ and its characteristic width $d$.

Due to electric neutrality the number of accumulated electrons has to compensate the external field $D_0$, i.e.
\begin{equation}
  \label{eq:neutrality}
  4 \pi e \int \limits_0^{\infty} n(x) dx = D_0.
\end{equation}
To take into account the electron screening of the external field we use the Thomas-Fermi approach \cite{Thomas_Fermi, Landau} in which the electron concentration $n(x)$ and self-consistent potential profile $\varphi(x)$ are related as $-e\varphi(x)+\mu(x)=\varepsilon_F=0$, where
\begin{equation}
  \label{eq:TF}
  \mu(x)= (3\pi^2)^{2/3}\frac{\hbar^2}{2 m} [n(x)]^{2/3}
\end{equation}
is the chemical potential of the electron gas. Thus, one can obtain the solution of Eqs. \eqref{eq:Gauss} by replacing  $\rho(x)$ with $e n(x)$ and using relations \eqref{eq:electric_field_small}, \eqref{eq:electric_field_high}. For a linear dielectric response we obtain the equation for the potential:
\begin{equation}
  \label{eq:potential_equation_linear}
\frac{d^2 }{dx^2} \left(\frac{\varphi}{e/b}\right) = \frac{2^{3/2}}{3\pi^2} \frac{1}{b^2} \frac{1}{\kappa} \left(\frac{\varphi}{e/b}\right)^{3/2}.
\end{equation}
We use the boundary condition $\varphi=0$  at $x \rightarrow \infty$ and get the solution \cite{Frenkel}:
\begin{equation}
  \label{eq:potential_linear}
  \varphi(x)=  C_1 \frac{e}{b} \kappa^2   \left(\frac{b}{x+d} \right)^4,
\end{equation}
\begin{equation}
  \label{eq:conventration_linear}
  n(x)= C_2 \frac{1}{b^3}   \kappa^3 \left(\frac{b}{x+d}\right)^{6},
\end{equation}
where  $C_1=(225/8)\pi^2 \simeq 278$, $C_2=1125 \pi /8 \simeq 442 $. For a nonlinear dielectric response we obtain the equation for the potential:
\begin{equation}
  \label{eq:potential_nonlinear_equation}
\frac{d}{dx} \left[\left(\frac{d}{dx} \frac{\varphi}{e/b}\right)^{1/3}\right] = \frac{2^{3/2}}{3\pi^2} \frac{1}{b^{4/3}} A^{1/3}\left(\frac{e/b^2}{P_0}\right)^{2/3} \left(\frac{\varphi}{e/b}\right)^{3/2}.
\end{equation}
With the same boundary condition we get the solution:
\begin{equation}
  \label{eq:potential_nonlinear}
  \varphi(x)=  C_3 \frac{e}{b} \left(\frac{b}{a}\right)^{8/7} \frac{1}{A^{2/7}} \left(\frac{b}{x+d} \right)^{8/7},
\end{equation}
\begin{equation}
  \label{eq:conventration_nonlinear}
  n(x)=C_4 \frac{1}{b^3} \left(\frac{b}{a}\right)^{12/7} \frac{1}{A^{3/7}}  \left(\frac{b}{x+d}\right)^{12/7},
\end{equation}
where $C_3= [5^6 3^6 \pi^{12}/(7^8 2^3)]^{1/7} \simeq 5.8,$ $C_4= [5^{9} 3^{2} \pi^{4} 2^6/7^{12}]^{1/7} \simeq 1.3.$

The characteristic length $d$ can be obtained using the neutrality condition (see Eq. (\ref{eq:neutrality})). For a linear dielectric response this gives:

\begin{equation}
  \label{eq:d_small}
  d= C_5 b \left(\frac{a}{b}\right)^{2/5} \kappa^{3/5} \left(\frac{e/a^2}{D_0}\right)^{1/5},
\end{equation}
where $C_5=[\pi^2 225/2 ]^{1/5} \simeq 4$. For a nonlinear dielectric response:

\begin{equation}
  \label{eq:d_STO}
 d= C_6 b \left(\frac{a}{b}\right)^{2/5} \left (\frac{e/a^2}{D_0}\right)^{7/5} \frac{1}{A^{3/5}},
\end{equation}
where $C_6 = (16/7) (5^2 3^2 \pi^{11})^{1/5} \simeq 84$. The induction $D_c$ at which the transition from linear to nonlinear dielectric response  occurs can be found from equating Eqs.  \eqref{eq:d_small} and   \eqref{eq:d_STO}. This gives
\begin{equation}
  \label{eq:4}
D_c = \frac{C_7}{\sqrt{A}} \frac{e}{a^2}\sqrt{\frac{1}{\kappa}},
\end{equation}
where $C_7=(2^{21} \pi^9/7^5)^{1/6} \simeq 12$, consistent with Eq. (\ref{eq:Dc}). For STO, the critical field $D_c$ depends on temperature:  $D_c \simeq 0.1 e/a^2$ for helium temperature and $D_c \simeq 0.7 e/a^2$ for room temperature.

The three dimensional concentration profile  $n(x)$ for the nonlinear dielectric response  Eq. (\ref{eq:conventration_nonlinear}) is the main result of this section. Note that $n(x)$ has a very long tail with a weak  $12/7$ power law dependence, which may lead to some arbitrariness in measurements of the width of the electron gas. Indeed, only 39\% of electrons are located within the distance $0< x < d$ near the interface and 68\% of electrons are located within $0<x < 4 d$. In the calculation above we used a continuous model. Actually,  along the [100] axis STO is composed of alternating $\mathrm{TiO_2}$ and $\mathrm{SrO}$ layers. The conduction band of $\mathrm{SrTiO_3}$ corresponds to the bands composed of mainly $3d$ orbitals of $\mathrm{Ti}$. Integrating $n(x)$ over each lattice cell in Table  \ref{tbl:distribution} we get a percentage of electrons in each of the 10 first $\mathrm{TiO_2}$ layers of STO for the case $D=2\pi e/a^2$.

\begin{table}%[H] add [H] placement to break table across pages
\caption{\label{tbl:distribution} Percentage of electrons   in the $\mathrm{TiO_2}$ layer $\textnumero M$ of STO for  $D_0=2\pi e/a^2$, corresponding to a total surface density of $0.5 e /a^2$ }
\begin{ruledtabular}
\begin{tabular}{ c   c c c c c c c c c c}
M & 1 & 2& 3&4&5&6&7&8&9&10 \\ \hline
\%&27.9&14.4&9.0&6.2&4.6&3.5&2.8&2.3&1.9&1.6 \\
%sdf&0.150 &0.074&0.045 &0.031 &0.022 &0.017 &0.014 &0.011 &0.009&0.008 \\
\end{tabular}
\end{ruledtabular}
\end{table}

One can see from Eqs. (\ref{eq:conventration_linear}) and (\ref{eq:conventration_nonlinear}) that the tails of the electron depth profiles $n(x)$ at $x \gg d$ do not depend on $D_0$ and behave like
\begin{equation}
\label{eq:x}
n(x)=C_2 \frac{1}{b^{3}} \kappa^3 \left(\frac{b}{x}\right)^6,
\end{equation}
and
\begin{equation}
\label{eq:y}
n(x)=C_4 \frac{1}{b^3} \left(\frac{b}{a} \right)^{12/7} \frac{1}{A^{3/7}} \left(\frac{b}{x}\right)^{12/7}
\end{equation}
for linear and nonlinear dielectric responses, respectively. Even for $D_0 \gg D_c$  when the electron distribution $n(x)$ at moderately large  $x$ is described by dependence (\ref{eq:conventration_nonlinear}), at very large distances the polarization becomes smaller and the linear dielectric response takes over so that the $n(x)$ dependence switches from Eq. (\ref{eq:conventration_nonlinear}) to Eq. (\ref{eq:conventration_linear}). This happens at the distance
\begin{equation}
\label{eq:z}
x_0=b \left(\frac{C_2}{C_4}\right)^{7/30}  A^{1/10} \left(\frac{a}{b}\right)^{2/5} \kappa^{7/10}
\end{equation}
($x_0 =360~\mathrm{nm}$ and $20~\mathrm{nm}$ for helium and room temperature respectively). Thus, the tail of $n(x)$ is universal. For small $D_0<D_c$ the tail has the form $n(x) \propto x^{-6}$. For $D_0>D_c$ it has the form  $n(x) \propto x^{-12/7}$ for $x<x_0$ and $n(x) \propto x^{-6}$ for $x>x_0$.

On the other hand, one has to remember  that  our theory is correct only when $n(x)$ is larger than the concentration of donors in the bulk of the material.

Let us verify whether the Thomas-Fermi approximation is applicable at $x\sim d$, i.e., $k_F d \gg 1$. Here $k_F = (3\pi^2)^{1/3}n(0)^{1/3}$ is the wavevector of an electron at the Fermi level. For $D_0 \ll D_c$

\begin{equation}
  \label{eq:linear_kfd}
  k_Fd = C_8 \kappa^{2/5} \left(\frac{b}{a}\right)^{2/5} \left(\frac{D_0}{e/a^2}\right)^{1/5},
\end{equation}
while  for $D_0 \gg D_c$
\begin{equation}
  \label{eq:cubic_kfd}
  k_Fd = C_9 \frac{1}{A^{2/5}} \left(\frac{b}{a}\right)^{2/5}  \left(\frac{e/a^2}{D_0}\right)^{3/5},
\end{equation}
where $C_8= (5^33^3\pi^3/2^4)^{1/5} \simeq 6$, $C_9=4/7 (15 \pi^3)^{3/5} \simeq 23.$ One can see that  $k_F d>1$ in the range of $2 \cdot 10^{-7} e/a^2 < D_0< 40 ~e/a^2 $ for room temperature. For lower temperatures this interval is even  larger. Thus, the Thomas-Fermi (TF) approximation is applicable for practically  all  reasonable induction $D_0$.

Let us now consider how the applicability of the TF approximation holds with growing $x$. The TF parameter $n^{1/3}x$ can be estimated with the help of Eqs. (\ref{eq:x}) and (\ref{eq:y}) and we then get
\begin{equation}
n^{1/3}x\simeq \left\{ \begin{array}{ll}
\left(\displaystyle{\frac{x}{b}}\right)^{3/7} , & x<x_0\,,\\
&\\
\left(\displaystyle{\frac{\kappa b}{x}}\right) \,\,\,,& x>x_0\,.
\end{array} \right.
\end{equation}

Thus, the TF approximation is valid in the whole range of nonlinear and linear dielectric response and fails only at very large distance from the interface $x=\kappa b$.
  \begin{comment}\footnote{So far we considered only two limiting cases: the linear and nonlinear dielectric responses, which are correct for $D_0 \ll D_c$ and $D_0 \gg D_c$ respectively. In fact one can analytically solve Eqs. (\ref{eq:Gauss}) with  Eq. (\ref{eq:electric_field_definition}) assuming that $D = 4\pi P$. For example, in Ref. \onlinecite{Gureev_ferroelectric_electrons}  the polarization profile of a charged ``head-to-head''  domain wall in a ferroelectric  was calculated. In the ferroelectric phase, the spontaneous polarization in the  domain is approximately equal to our $D_c$. Therefore a charged domain wall,  considered in Ref. \onlinecite{Gureev_ferroelectric_electrons}, should be compared with our accumulation layer at the crossover value of the electric field, $D_0=D_c$. Indeed, our results Eqs. (\ref{eq:conventration_linear}), (\ref{eq:d_small}) taken at $D_0=D_c$  agree with those for the charged domain wall.}\end{comment}

\section{Two overlapping accumulation layers}
\label{sec:overlap}
\begin{figure}[h]
\includegraphics[width=.5\textwidth]{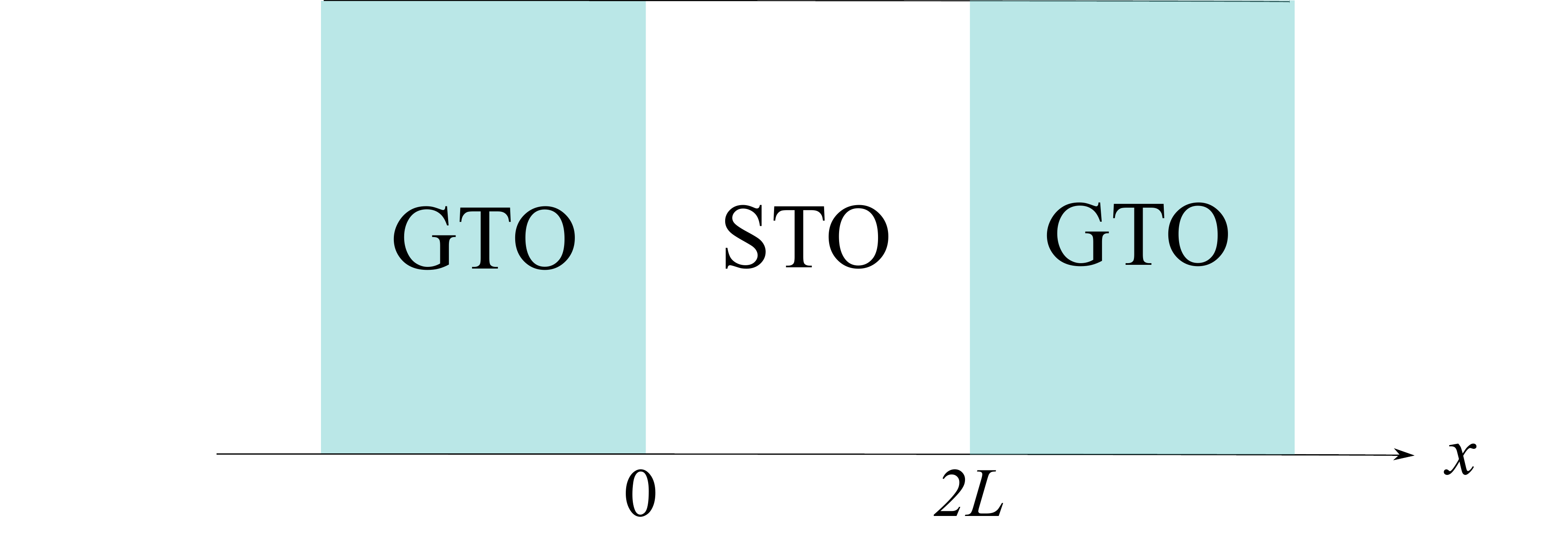}\\
\caption{GTO/STO/GTO structure with wide enough GTO layers and an STO layer of width $2L$. The GTO layers generate $D_0=2\pi e/a^2$ on each interface.}\label{fig:2}
\end{figure}

In the  section \ref{sec:singlep}, we have investigated a single accumulation layer induced in STO. In the GTO/STO/GTO structure \cite{Stemmer_2012, Stemmer_2013, Stemmer_2015}, an accumulation layer forms near each interface and these two layers overlap with each other. When the STO layer is thick, one can expect that the two accumulation layers overlap weakly by the vanishing tails and the final electron distribution can be described as a simple addition of two accumulation layers given by Eq. (\ref{eq:conventration_nonlinear}). (The induction $D_0$ on each interface is $4\pi e/2a^2>D_c$ and we are interested only in the region where the distance to the interface is smaller than $x_0$. So we use the nonlinear dielectric response here.) However, as the STO layer gets thinner, the overlap becomes stronger. Due to the nonlinear physics here, the electron density profile changes substantially. Below we study the density profile of the electron gas in the GTO/STO/GTO structure where the width of the STO layer is $2L$ (see Fig. \ref{fig:2}).

To make the mathematics more compact, in Secs. \ref{sec:overlap} and \ref{sec:spillout}, we employ the dimensionless notations, in which Eq. (\ref{eq:potential_nonlinear_equation}) is rewritten as
\begin{equation}
\frac{d}{d \xi}\left(\frac{d\chi}{d \xi}\right)^{1/3}=\alpha\chi^{3/2},\quad \xi>0.\label{eq:20}
\end{equation}
Here $\xi=x/b$ is the scaled distance from one interface, $\chi=\varphi/(e/b)$ is the potential in the units of $e/b$, and $\alpha=2^{3/2}(a/b)^{4/3}A^{1/3}/3\pi^2$. Also, from Eq. (\ref{eq:TF}), we know
\begin{equation}
n(x)=\frac{C_{10}}{b^3}\left[\frac{\varphi(x)}{e/b}\right]^{3/2},
\label{eq:5n}
\end{equation}
where $C_{10}=2^{3/2}/3\pi^2\approx 0.1$. The electron density $n(x)$ can be scaled as $\widetilde{n}(\xi)=A^{1/3}(a/b)^{4/3}n(x)b^3$ and we then get
\begin{equation}
\label{eq:new}
\widetilde{n}(\xi)=\alpha\chi^{3/2},
\end{equation}
which is the dimensionless form of the TF approximation.
The width of the STO layer is scaled as $2\widetilde{L}=2L/b$.

According to Eq. (\ref{eq:20}), we can get
\begin{equation}
\frac{d\chi}{d\xi}=-\left(\frac{8}{5}\alpha\chi^{5/2}+g_1\right)^{3/4}\label{eq:21},
\end{equation}
where $g_1$ is a constant arising from the integration. By
integrating Eq. (\ref{eq:21}), we get
\begin{equation}
\int_{\chi(\widetilde{L})}^{\chi(0)}\frac{d\chi}{\left(\frac{8}{5}\alpha\chi^{5/2}+g_1\right)^{3/4}}=\int_0^{\widetilde{L}} d\xi=\widetilde{L}.\label{eq:22}
\end{equation}
In this structure, each GTO/STO heterojunction provides a fixed number of electrons to the accumulation layer inside STO which is $e/2$ per unit cell with the interfacial induction $D_0=4\pi e/2a^2$. Using Eq. (\ref{eq:electric_field_high}), this gives the value of $d\chi/d\xi$ on the $\xi=0$ and $\xi=2\widetilde{L}$ interfaces,  which is $\propto E\propto D_0^3$. Due to the symmetry, the electric field is zero in the middle of the STO layer which means $d\chi/d\xi=0$ at $\xi=\widetilde{L}$. We can choose a value for $g_1$ and calculate $\chi(\widetilde{L})$ and $\chi(0)$ using Eq. (\ref{eq:21}). Then we can put boundary values of $\chi$ into Eq. (\ref{eq:22}) and get the corresponding value of $\widetilde{L}(g_1)$. Reversing $\widetilde{L}(g_1)$, we find the function $g_1(\widetilde{L})$. Therefore, at any given value of $\widetilde{L}$ we can pin down $g_1$ and use Eq. (\ref{eq:21}) to numerically get the whole electron profile inside STO.

\begin{figure}[h]
$\begin{array}{c}
\includegraphics[width=0.5\textwidth]{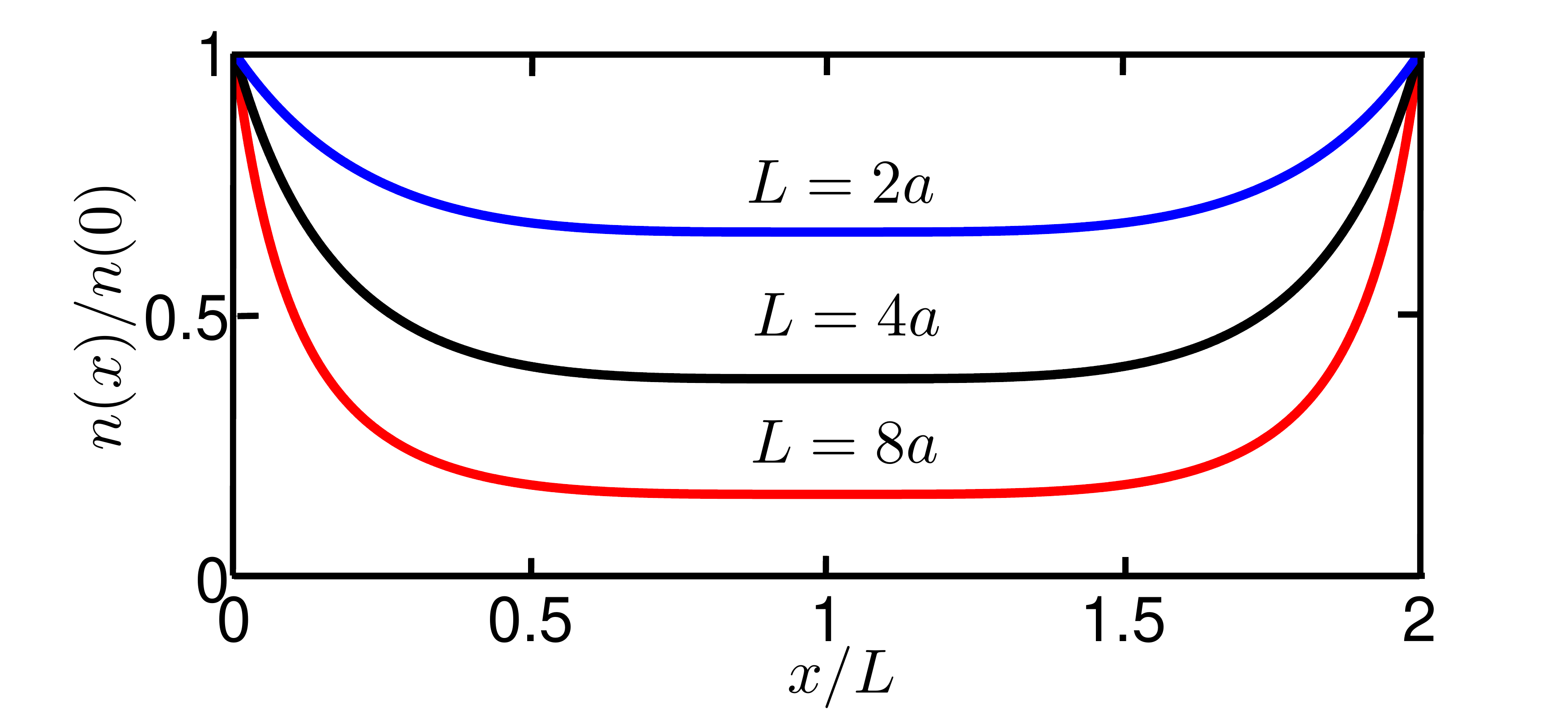}\\
\end{array}$
\caption{(Color online) Electron distribution at different values of $L$ which is the half-width of the STO layer ($a$ is the lattice constant). The thick solid lines are the electron concentration $n(x)$ in the unit of $n(0)$ where $n(0)$ is the electron concentration at $x=0$ (red: $L=8a$; black: $L=4a$; blue: $L=2a$). When $L$ is not very large, the electron density profile is substantially different from the addition of two accumulation layers described by Eq. (\ref{eq:conventration_nonlinear}).}\label{fig:5}
\end{figure}

To realize this, we need to try various values of $g_1$ and tune accordingly until we find the $\widetilde{L}$ we want. In this process, it is necessary to know what values of $g_1$ are physically possible. It is obvious that the extreme values of $g_1$ appear at $\widetilde{L}\rightarrow 0$ and $\widetilde{L}\rightarrow \infty$. For the former case, electrons are almost uniformly distributed over the thin STO layer with $d\chi/d\xi\simeq 0$ and the electron density is approximately $1/2a^2\widetilde{L}$ everywhere which gives the value of $\chi$ inside STO by Eq. (\ref{eq:new}) (see Fig. \ref{fig:5}). From Eq. (\ref{eq:21}), we know at $d\chi/d\xi=0$, a bigger value of $\chi$ corresponds to a smaller value of $g_1$. Thus, in this case we get the minimum value of $g_1$ as
\begin{equation}
{g_1}^{min}=-\frac{8\left[A^{1/3}(a/b)^{4/3}b^3(0.5/a^2\widetilde{L})\right]^{5/3}}{5\alpha^{2/3}}.\label{eq:23}
\end{equation}
Since $\widetilde{L}$ can be arbitrarily small (the lattice constant $a$ is regarded as infinitesimal), we get in fact ${g_1}^{min}=-\infty$.

In the latter case where the STO layer is very thick, the two interfaces are quite independent and tails of accumulation layers barely overlap. So, near the center of the STO layer where $d\chi/d\xi=0$, the electron density vanishes. This gives the maximum value of $g_1$ as ${g_1}^{max}=0$.
In this case, the electron distribution is close to the simple addition of two accumulations layers described by Eq. (\ref{eq:conventration_nonlinear}) (see Fig. \ref{fig:5}).

So we get the domain of $g_1$ as $(-\infty,\,0\,]$. This means we can choose whatever value for $g_1$ and get physically meaningful results. By trying different values of $g_1$, we can find the electron density profiles at certain $\widetilde{L}$ that we are interested in. In Fig. \ref{fig:5}, we show our results for electron density profiles at 3 different values of $L$. We see the evolution from an almost constant $n(x)$ at $L=2a$ around which the quantum criticality is observed \cite{Stemmer_2015} to the one reminiscent of two weakly overlapping tails of accumulation layers described by Eq. (\ref{eq:conventration_nonlinear}) at $L=8a$.

It is easy to see that $n(x)$ is quite flat near the middle. This is actually an effect of the nonlinear dielectric response as we show below.
Indeed, due to the symmetry of the structure, for the potential $\chi(\xi)$, derivatives of odd order including $d\chi/d\xi$ are all vanishing in the middle. Using Eq. (\ref{eq:20}), the second-order derivative $d^2\chi/d\xi^2$ is found to be proportional to $d\chi/d\xi$ and is then also zero at $\xi=\widetilde{L}$ while the fourth-order is nonzero.
\begin{comment}
\begin{equation}
\frac{d^2\chi}{d\xi^2}=3\alpha\chi^{3/2}\left(\frac{d\chi}{d\xi}\right)^{2/3},
\end{equation}
while the fourth-order term is
\begin{equation}
\frac{d^4\chi}{d\xi^4}=\frac{9\alpha}{4\chi^{1/2}}\left(\frac{d\chi}{d\xi}\right)^{8/3}+\frac{81\alpha^2\chi^2}{2}\left(\frac{d\chi}{d\xi}\right)^{4/3}+6\alpha^3\chi^{9/2}.
\end{equation}\end{comment}
This explains why the density changes so slowly near the middle of the STO layer. We emphasize again that this ``flattening" effect originates from the nonlinear dielectric response. When the response is linear, the differential equation is described by Eq. (\ref{eq:potential_equation_linear}), according to which the second-order derivative $d^2\chi/d\xi^2$ is nonzero even at $d\chi/d\xi=0$. The density change near the middle is then faster.

\section{Spill-out of electrons from heavily doped $n$-type STO (NSTO) into STO}
\label{sec:spillout}

\begin{figure}[h]
\includegraphics[width=.5\textwidth]{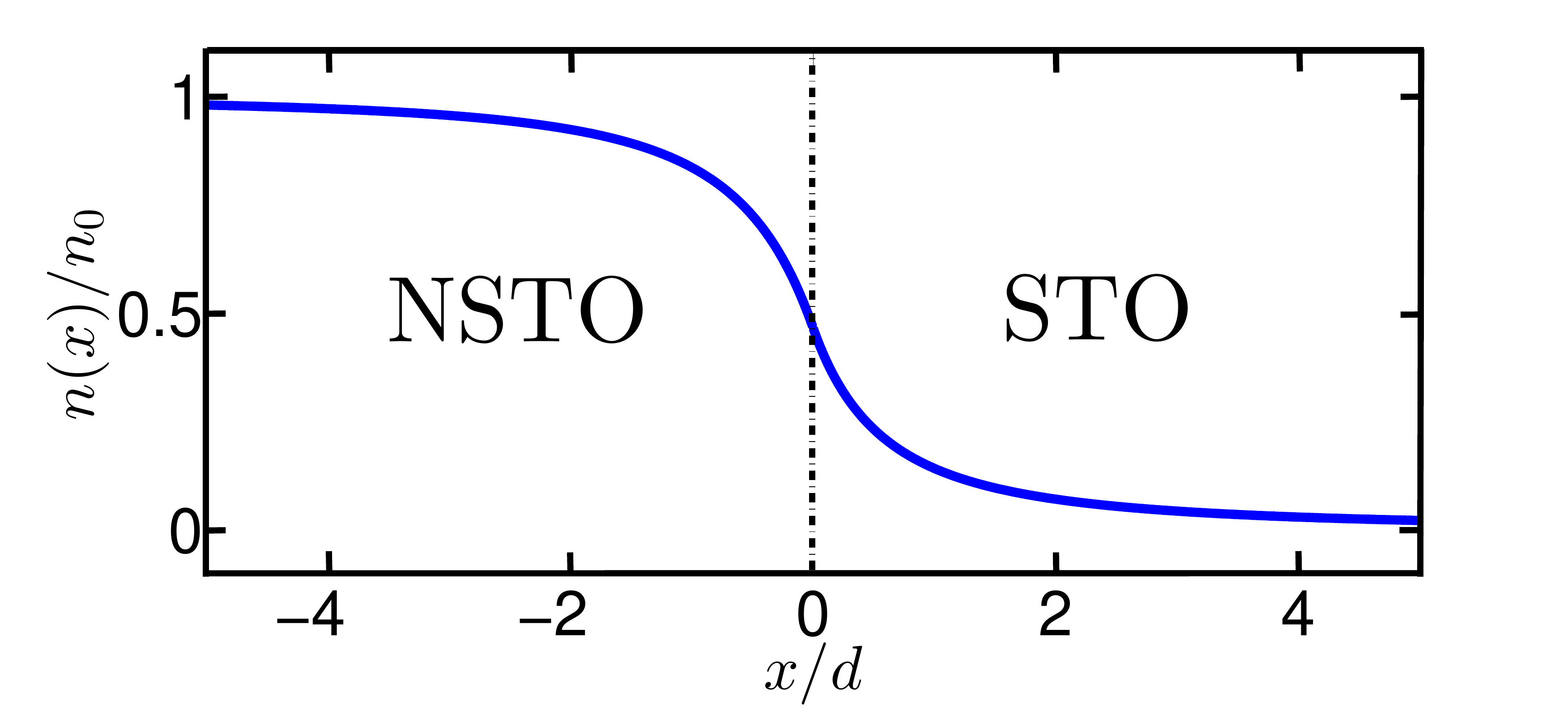}\\
\caption{The electron distribution in a single NSTO/STO junction. The distance $x$ is in units of the decay length $d=b\widetilde{d}$. The electron density $n(x)$ is in units of the donor concentration $n_0$. This figure is universal for all values of $n_0$.}\label{fig:4}
\end{figure}

In the NSTO/STO junctions \cite{Hwang_2009, Stemmer_2010}, the interface is formed between a moderately $n$-type doped STO with the Fermi level close to the conduction band bottom (below we simply denote it as STO) and a heavily $n$-type doped STO (NSTO), which has a much higher Fermi level. We start from considering a single junction between thick layers of NSTO and STO. In this case as a result of the original difference between Fermi levels electrons spill out into STO to create a common Fermi level and the total number of spilled electrons depends only on the doping level inside NSTO. Below, we study the electron distribution for this scenario  (see Fig. \ref{fig:4}) assuming that the donor concentration in NSTO is so large that dielectric response is nonlinear. (The linear case of such spill-out problems was first addressed by Frenkel \cite{Frenkel}.)

Inside NSTO, the charge concentration is $[n_0-n(x)]e$ where $n_0$ is the density of the positive background charge inside the doping layer and $n(x)$ is the electron density at distance $x$ from the interface. According to the Thomas-Fermi approximation, we have
\begin{equation}
\frac{d}{d\xi}\left(\frac{d \chi}{d\xi}\right)^{1/3}=\alpha\chi^{3/2}-\widetilde{n}_0,\quad \xi\le 0.\label{eq:26}
\end{equation}
Here $\widetilde{n}_0=A^{1/3}(a/b)^{4/3}n_0b^3$ is the scaled background charge concentration.
From Eq. (\ref{eq:26}), it can be derived that
\begin{equation}
\begin{aligned}
\frac{d\chi}{d\xi}=&-\left(\frac{8}{5}\alpha\chi^{5/2}-4\widetilde{n}_0\chi+g_2\right)^{3/4},\quad\xi\le0\label{eq:27}
\end{aligned}
\end{equation}
where $g_2$ is a constant that we determine from the boundary conditions. Since at $\xi\rightarrow-\infty$, $\widetilde{n}(\xi)=\widetilde{n}_0$, we have $d\chi/d\xi=0$ and $\chi=\left(\widetilde{n}_0/\alpha\right)^{2/3}$. Therefore,
\begin{equation}
g_2=\frac{12}{5}\frac{{\widetilde{n}_0}^{\,5/3}}{\alpha^{2/3}}.\label{eq:28}
\end{equation}

Now let us switch to the STO side.
By rewriting Eq. (\ref{eq:potential_nonlinear}) in the dimensionless form, we have in STO
\begin{equation}
\chi=\frac{\chi_c}{(\xi+\widetilde{d})^{\,8/7}},\quad\xi>0\label{eq:29}
\end{equation}
which gives
\begin{equation}
\label{eq:vanish_g_1}
\frac{d\chi}{d\xi}=-\left(\frac{8}{5}\alpha\chi^{5/2}\right)^{3/4}.
\end{equation}
Here $\widetilde{d}=d/b$ is the scaled decay length and $\chi_c=\left(2^35^3/7^4\alpha^3\right)^{2/7}$.
Since the interfacial field $D_0$ is now no longer a fixed value, we cannot use Eq. (\ref{eq:d_STO}) to get the decay length. Instead, using the boundary condition that $\chi$ and $d\chi/d\xi$ are continuous at $\xi=0$ which satisfy both Eqs. (\ref{eq:27}) and (\ref{eq:vanish_g_1}), we then have
\begin{equation}
g_2=4\widetilde{n}_0\chi(0)=\frac{4\widetilde{n}_0\chi_c}{\widetilde{d}^{\,\,8/7}}\label{eq:30}.
\end{equation}
Together with Eq. (\ref{eq:28}), we then have
\begin{equation}
\widetilde{d}=\left(\frac{5\chi_c}{3}\right)^{7/8}\left(\frac{\alpha}{\widetilde{n}_0}\right)^{7/12},\label{eq:32}
\end{equation}
which gives the expression for the decay length.

Now we get the electron distribution inside STO. Using the differential equation (\ref{eq:27}) and the boundary values of $\chi$ and $d\chi/d\xi$ at $\xi=0$, the electron density profile inside NSTO can also be obtained numerically. A schematic plot of the electron distribution is presented in Fig. \ref{fig:4}. This is the universal curve independent of the value of $\widetilde{n}_0$.

Let us now dwell on the case of a heavily doped NSTO layer with the width $2t$ embedded in STO (STO/NSTO/STO structure). Experimentally used  $\delta$ layers have sharp boundaries \cite{mobility_sharp_interface_STO}  with well defined volume concentration $n_0$. One can imagine two limiting cases. When NSTO layer is so wide that its half-width $t$ is much larger than the width of the electron gas in STO, one can use the above single-junction theory for the two separate STO/NSTO and NSTO/STO junctions.

In the opposite case when $t$ is much smaller than the width $d$ of the electron gas on each side, most of electrons are located outside of such $\delta$-layer. As a result the effective two-dimensional charge density $2tn_0$ creates two accumulation electron gases on both sides from the $\delta$-layer. The external electric field for each gas is $D_0=4\pi n_0 t$. The width of the gas is determined by Eq. \eqref{eq:d_STO} if $D_0 > D_c$ which is given by Eq. \eqref{eq:4}.

Below we discuss how such peculiar distribution of electrons may affect the low temperature transport properties of the structure. Namely, for a fixed concentration $n_0$ we find the ratio of the mobility $\mu_{\delta}$ for a $\delta$-doped structure with $t \ll d$  to the bulk mobility $\mu_{b}$ for a structure with  $t \gg d$.

We focus on low temperatures where electrons are mostly scattered by donors and the phonon scattering is negligible.  The scattering crossection of an electron on an ionized donor is $\Sigma \simeq a^2$ ($a$ is the lattice constant) and does not depend on the electron energy. Indeed, the Coulomb impact parameter for an electron with Fermi energy $\varepsilon_F=(3\pi^2 n)^{2/3} \hbar^2/2m$ is  $I=e^2/\kappa \varepsilon_F$. For concentration $n$ larger than $10^{17}~\mathrm{cm^{-3}}$, $I \ll a$. As a result the cross-section saturates at the neutral-impurity cross-section $\Sigma \simeq a^2$. This unique feature of materials with high dielectric constant \cite{Wemple_KTaO_transport} simplifies the calculation of the mobility. In the case of $\delta$-doping, most of electrons live outside of the layer and are scattered by donors only when they move through this layer. This leads to a reduced effective concentration of scattering centers $n_0t/d$ which is much smaller than the concentration $n_0$ in the $t\gg d $ case. At $t \ll d$ the Fermi velocity of the electron gas is $v_{\delta} \simeq (\hbar/m) (n_0t/d)^{1/3} \propto t^{4/5}$ since $d\propto t^{-7/5}$ for the nonlinear dielectric response, while for $t \gg d$ the velocity $v_{b}$ does not depend on $t$. In the result,  $\mu_{\delta}/\mu_{b}=(d/t) (v_b/v_{\delta}) \propto t^{-16/5}$.  Thus, the mobility $\mu_{\delta}$ decreases with  $t$ at $t \ll d$ and at $t \simeq d$ saturates as the bulk value $\mu_b$. This effect probably was observed in Ref.   \onlinecite{mobility_delta_doped_STO}.

\begin{figure}[h]
\includegraphics[width=.5\textwidth]{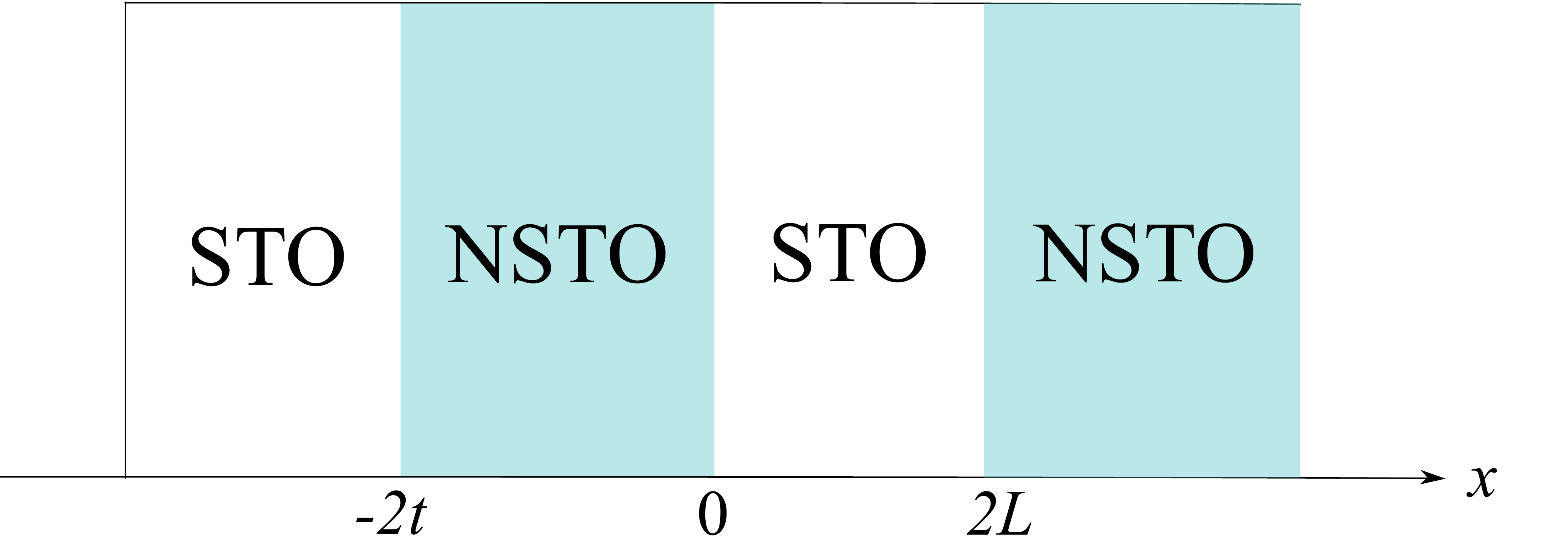}\\
\caption{Periodic NSTO/STO structure. The width of the NSTO layer is $2t$ and the width of the STO layer is $2L$.}\label{fig:3}
\end{figure}

\begin{figure}[h]
$\begin{array}{c}
\includegraphics[width=0.5\textwidth]{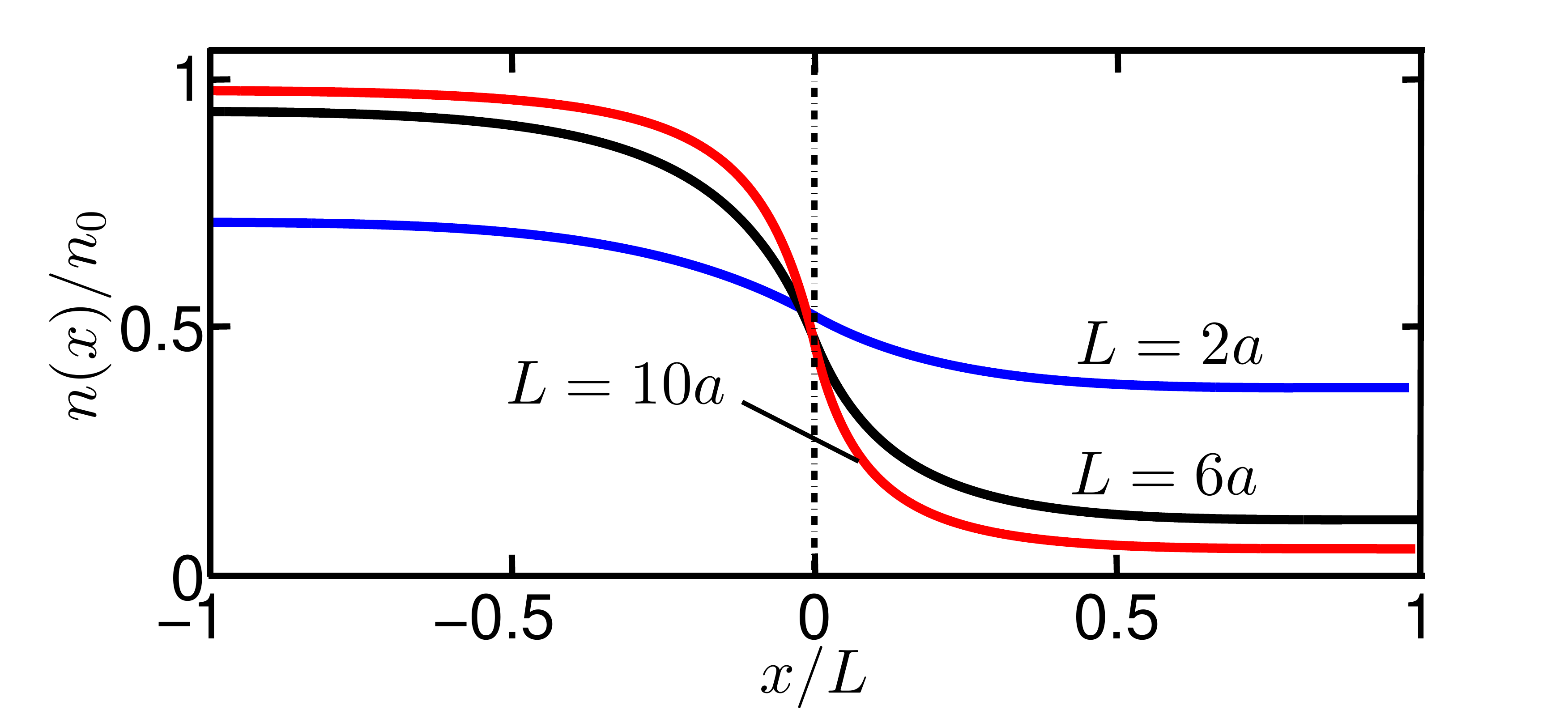}\\
\end{array}$
\caption{(Color online) Electron distribution from $-t$ to $L$ plotted at $t=L$ for different values of $L$. The thick solid lines are the electron density $n(x)$ in the unit of $n_0$ where $n_0$ is the donor concentration in NSTO (red: $L=10a$; black: $L=6a$; blue: $L=2a$). This graph is plotted at $n_0=0.5/a^3$ where $a$ is the lattice constant. As the layers get thinner, the electron distribution gets more uniform, deviating from the one shown in Fig. \ref{fig:4}.}\label{fig:6}
\end{figure}

Let us now address more complicated periodic structures formed by NSTO and STO \cite{Ohtomo_2002,Chang_2013,Choi_2012} (see Fig. \ref{fig:3}).
Inside each NSTO layer, the potential should satisfy Eq. (\ref{eq:27}) while in STO, it obeys Eq. (\ref{eq:21}). Similar to Eq. (\ref{eq:22}), we get
\begin{equation}
\int_{\chi(0)}^{\chi(-\widetilde{t})}\frac{d\chi}{\left(\frac{8}{5}\alpha\chi^{5/2}-4\widetilde{n}_0\chi+g_2\right)^{3/4}}=\int_{-\widetilde{t}}^0 d\xi=\widetilde{t}.\label{eq:31}
\end{equation}
where $\widetilde{t}=t/b$  and $2\widetilde{t}$ is the dimensionless width of the NSTO layer.
Similarly to what we have done above, we can choose certain values for $g_2$ and $\chi(0)=\chi_0$ where $\chi(0)$ is the scaled potential on the interface. Then we can calculate corresponding $\widetilde{t}$ and $\widetilde{L}$ and the electron density profile. Thus, in reverse, at any given $\widetilde{t}$ and $\widetilde{L}$ we can find the electron distribution.
Again, we need to find the physical range of $g_2$ and $\chi_0$, which can be done similarly to what we did in Sec. \ref{sec:overlap}.
Given values of $\widetilde{t}$ and $\widetilde{L}$, we try different $g_2$ and $\chi_0$ from their domains until we find the $\widetilde{t}$ and $\widetilde{L}$ we want. In this way, we get the electron distribution for 3 different values of $\widetilde{L}$ in Fig.
 \ref{fig:6} (for simplicity, we choose $\widetilde{t}=\widetilde{L}$ here). We see that as $L=b\widetilde{L}$ increases, the periodic electron density profile $n(x)$ evolves from being relatively constant to strongly oscillating. Again, one can see a ``flattening" effect near the middle of each layer, either STO or NSTO. Like what we did in Sec. \ref{sec:overlap}, one can verify that all derivatives of $\chi$ vanish in the middle of both layers until the fourth-order one.

Let us also dwell on a recently discovered new type of STO-based heterojunction formed by STO and NdTiO$_3$ (NTO). In this structure, the ceiling of the valence band created by the Mott gap in NTO is above the bottom of the conduction band of STO \cite{Bharat_2014, Bharat_aps}. So in addition to the ``polar discontinuity" \cite{Ohtomo_2004, Reinle_2012, Chris}, the broken-gap band alignment further brings electrons into STO by the electron "spill-out" and the electron distribution becomes more complicated. One can basically use the methods we employed in this section to solve the problem. The difference due to the presence of the ``polar discontinuity" is the abrupt jump of the induction field on the interface which gains an additional value of $4\pi e/2a^2$ when going from NTO to STO. One should note that the dielectric response inside NTO is linear and the potential obeys Eq. (\ref{eq:potential_equation_linear}). Therefore one gets an expression of the potential derivative $d\chi/d\xi$ different from Eq. (\ref{eq:27}). Also one should know the density of states below the Mott gap in NTO in order to solve for the electron distribution. The rest of the procedure is quite similar and one can then obtain the electron density profile in this structure.

\section{Spherical Donor Clusters}
\label{sec:spherical}
So far, we have considered only the planar structures based on STO. In Secs. \ref{sec:spherical} and \ref{sec:cylindrical}, we extend our studies to other non-planar geometries (spherical and cylindrical). To make the physics clearer, we return to the dimensional expressions for physical quantities from now on as in Sec. \ref{sec:singlep}.

\subsection{Renormalization of Charge}
\begin{figure}[h]
\includegraphics[width=0.5\textwidth]{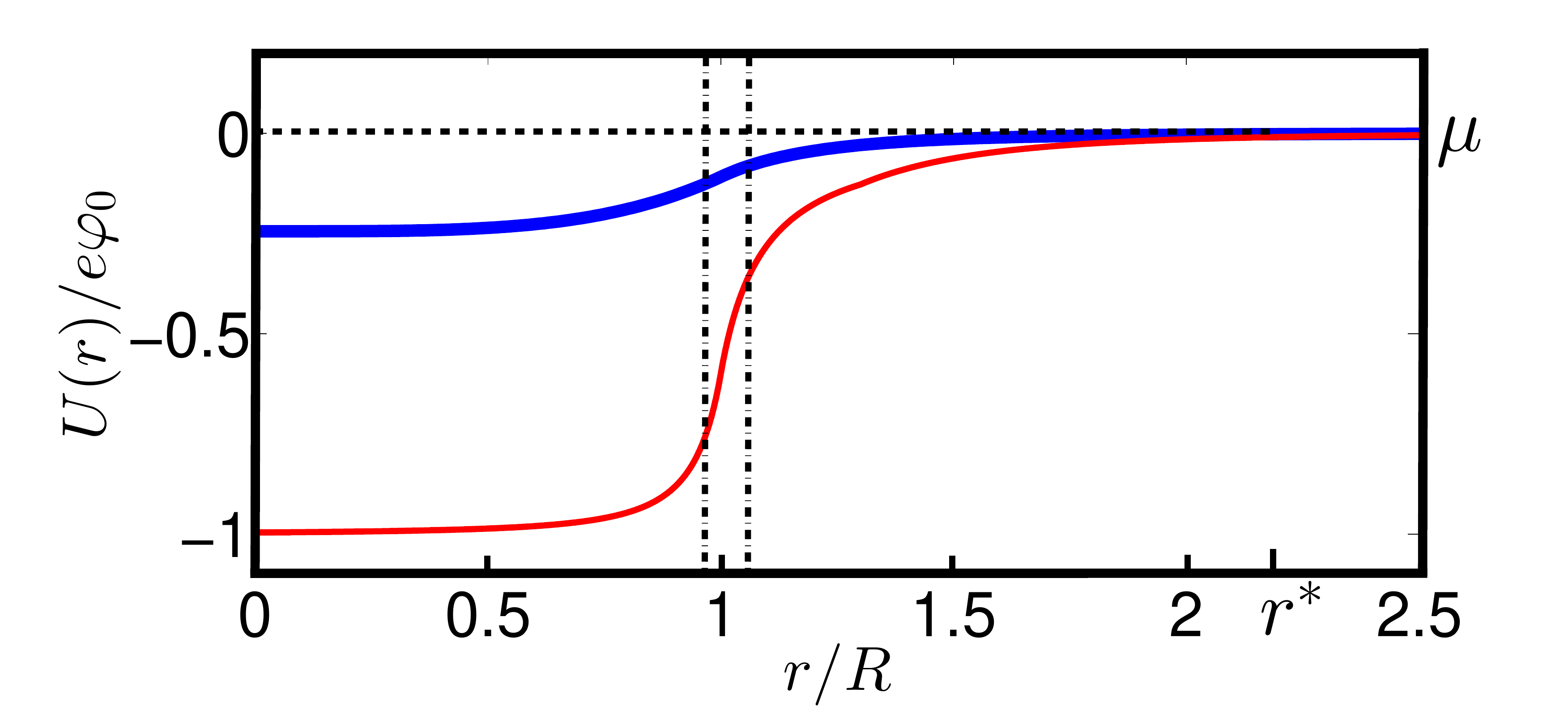}
\caption{(Color online) Potential energy of electrons $U(r)=-e\varphi(r)$ caused by the spherical donor cluster of radius $R$ as a function of distance $r$ from the cluster center. $\varphi_0$ is defined as $n(\varphi_0)=n_0$, where $n_0=3Z/4\pi R^3$, $n(r)$ is a function of $\varphi(r)$ given by Eq. (\ref{eq:5n}). The thick solid line (blue) represents the potential profile of a cluster of charge $Z\lesssim Z^*$ which is in the regime of weak charge renormalization. The thin solid line (red) represents the potential of a cluster at $Z\gg Z^*$ in the strong renormalization regime, where the two vertical dotted lines show edges of the ``double-layer" structure of width $\sim d\ll R$. The horizontal dashed line (black) indicates the position of the chemical potential $\mu= 0$. $r^*$ is the external radius of the collapsed electron gas where Thomas-Fermi approach fails.}
\label{fig:3n}
\end{figure}

Consider a large spherical donor cluster of the radius $R$ and the total positive charge $Ze$ such that $a\ll R< \kappa b/Z$ (for example, $R$ can be $3$ nm and $Z\simeq60$). If the dielectric response is linear, the electrons are mainly located at distances between $r_1=\kappa b/Z$ and $r_A=\kappa b$ from the cluster \cite{Landau}. For a very large $\kappa$, these radii are huge ($r_A=700$ nm in STO at liquid helium temperature) and the electrons are far away from the cluster. However, at small distances, the dielectric response is nonlinear and changes the potential form. If the potential energy outweighs the kinetic energy, electrons are attracted to the cluster and renormalize the net charge. To see when this happens, we look at the specific form of electric potential in this situation. We can calculate the potential from the differential equations (\ref{eq:9a}) and (\ref{eq:9b}) applicable to the spherical structure:
\begin{subequations}
\begin{align}
\left(\frac{d}{dr}+\frac{2}{r}\right)\left(\frac{d\varphi}{dr}\right)^{1/3}&=\frac{A^{1/3}e}{P_0^{2/3}}\left[n(r)-n_0\right],\, r<R\label{eq:9a}\\
\left(\frac{d}{dr}+\frac{2}{r}\right)\left(\frac{d\varphi}{dr}\right)^{1/3}&=\frac{A^{1/3}e}{P_0^{2/3}}n(r), \quad\quad\,\quad r>R\label{eq:9b}
\end{align}
\end{subequations}
where $r$ is the radius from the cluster center, $n(r)$ is the electron density at radius $r$ and $n_0$ is the donor concentration inside the cluster. However, due to the simple charge distribution here, we can get the potential in an easier way.  At $r>R$, the sphere looks like a point charge and $D(r)=Ze/r^2$. Using this together with Eqs. (\ref{eq:Gauss}) and (\ref{eq:electric_field_definition}), one can calculate the electric field
and get the electric potential $\varphi(r)$ as:
\begin{equation}
\varphi(r)=\frac{A}{P_0^2}\left(\frac{Ze}{4\pi}\right)^3\frac{1}{5r^5}, \quad\quad\quad\quad\quad\quad R<r\ll r_1
\label{eq:3b}
\end{equation}
with $\varphi(r=\infty)$ defined as zero. Inside the cluster at $r<R$, since the charge is uniformly distributed over the sphere, the total positive charge enclosed in the sphere of radius $x$ is equal to $Zer^3/R^3$, so $D(r)=Zer/R^3$. One then gets the corresponding potential $\varphi(r)$:
\begin{equation}
\varphi(r)=\frac{A}{P_0^2}\left(\frac{Ze}{4\pi}\right)^3\left(\frac{9}{20}\frac{1}{R^5}-\frac{1}{4}\frac{r^4}{R^9}\right), 0<r<R
\label{eq:3a}
\end{equation}
using the boundary condition $\varphi(r=R^-)=\varphi(r=R^+).$
A schematic graph of the potential energy $U(r)=-e\varphi(r)$ is shown in Fig. \ref{fig:3n} by the thick solid line (blue).

The Hamiltonian for a single electron is $H=p^2/2m^*-e\varphi(r)$, where $p$ is the momentum of the electron and $m^*$ is the effective electron mass in STO \cite{RS}. If we approximately set $p\simeq\hbar/2r$, we get a positive total energy of the electron everywhere when $Z$ is very small. This means there are no bound states of electron in the cluster. However, when $Z$ is big enough so that $Z>Z_c$, the electron can have negative total energy at $r<R$ and will collapse into the cluster.
Using Eqs. (\ref{eq:3b}) and (\ref{eq:3a}), we find
\begin{equation}
%r_0&=0.88R\\%\left(\frac{3}{5}\right)^{1/4}\\
Z_c\approx\frac{4\pi(b/Aa)^{1/3}R}{a}% \left[\frac{1}{4\left(3/5\right)^{3/2}}\right]^{1/3}
\sim \frac{R}{a}\gg 1.
\end{equation}
As $Z$ continues increasing, more and more electrons get inside the cluster filling it from the center where the potential energy is lowest (see Fig. \ref{fig:3n}). The single-electron picture no longer applies. Instead, we use the Thomas-Fermi approximation \cite{Landau} with the electrochemical potential $\mu=0$, which gives Eq. (\ref{eq:5n}). (We continue to assume here that the bulk STO is a moderately doped semiconductor.)

When the number of collapsed electrons $S$ is small, their influence on the electric potential is weak. One can still use Eqs. (\ref{eq:3b}) and (\ref{eq:3a}) for $\varphi(r)$ and get the corresponding expression of $n(r)$. At $r>R$, since $\varphi(r)$ is $\propto 1/r^5$, we get $n(r)\propto1/r^{15/2}$. In this way, we calculate $S$ as
\begin{equation}
S=\int^\infty_0 n(r)4\pi r^2dr=0.5Z\left(\frac{Z}{Z^*}\right)^{7/2}\propto Z^{9/2},
\label{eq:6n}
\end{equation}
where %the electrons collapsed onto the surface of the cluster instead of being inside accounts for $0.1(Z/Z^*)^{7/2}Z$ and
\begin{equation}
Z^*=\left[\frac{4\pi(b/Aa)^{1/3}R}{a}\right]^{9/7},
\label{eq:7n}
\end{equation}
The net charge number of the cluster is
\begin{equation}
Z_n=Z-S=Z\left[1-0.5\left(\frac{Z}{Z^*}\right)^{7/2}\right].
\label{eq:8n}
\end{equation}
One can see, when $Z_c\ll Z\ll Z^*$, one gets $S\ll Z$ and $Z_n\lesssim Z$, meaning the charge renormalization is weak. However, at $Z\sim Z^*$, according to Eqs. (\ref{eq:6n}) and (\ref{eq:8n}), we get $Z_n\sim S\sim Z^*$. The potential contributed by electrons is no longer perturbative. This brings us to the new regime of strong renormalization of charge.

$\phantom{}$

We show that at $Z\gg Z^*$ the net charge $Z_ne$ saturates at the level of $Z^*e$. Indeed, when $Z$ grows beyond $Z^*$, $Z_n$ can not go down and therefore can't be much smaller than $Z^*$.
At the same time it can not continue going up, otherwise as follows from Eqs. (\ref{eq:3b}) and (\ref{eq:6n}) with $Z$ replaced by $Z_n\gg Z^*$, the total electron charge surrounding the charge $Z_ne$ at $r > R$ would become $Se \simeq Z_ne (Z_n/Z^*)^{7/2}\gg Z_ne$ leading to a negative charge seen from infinity. Thus, at $Z\gg Z^*$, the net charge $Z_n$ saturates at the universal value of the order of $Z^*$. This result is qualitatively similar to the one obtained for heavy nuclei and donor clusters in Weyl semimetals and narrow-band gap semiconductors in Ref. \onlinecite{Kolomeisky}.

In the following subsection, we show how the renormalization of charge at $Z\gg Z^*$ is realized through certain distribution of electrons, in which a structure of ``double layer" (see Fig. \ref{fig:3n}) plays an important role.

\subsection{Radial Distribution of Electrons}

At $Z\gg Z^*$, the charge renormalization is strong and the most of the sphere of radius $R$ is completely neutralized by electrons. In the neutral center of the sphere, the electron density $n(r)=n_0$, where $n_0=3Z/4\pi R^3$ is the density of the positive charge inside the cluster. The corresponding ``internal" electric potential $\varphi_{in}(r)=\varphi_0$ where $\varphi_0$ is given by $n(\varphi_0)=n_0$ using Eq. (\ref{eq:5n}). $\varphi_{in}(r)$ is then $\propto (n_0a^3)^{2/3}\propto [Z/(R/a)^3]^{2/3}$. Outside the cluster, when the charge is renormalized to $Z_n$, one gets a potential $\varphi_{out}(r)$ similar to Eq. (\ref{eq:3b}) with $Z$ replaced by $Z_n$. Since $Z_n$ is $\sim Z^*$ where $Z^*$ is given by Eq. (\ref{eq:7n}), we get $\varphi_{out}(r)\propto (R/a)^{-8/7}$ at a distance $r$ of the order $R$. Thus, close to the cluster surface, the ratio of the outside potential $\varphi_{out}(r)$ to the inside potential $\varphi_{in}(r)$ is $\simeq(R/a)^{6/7}/Z^{2/3}\ll 1 $ since $Z\gg Z^*\simeq(R/a)^{9/7}$. This indicates a sharp potential drop across the sphere surface.

At $0<R-r\ll R$, there's a thin layer of uncompensated positive charges. At $0<r-R\ll R$, a higher potential than farther away means a larger electron concentration that forms a negative layer close to the surface. This ``double-layer" structure resembles a capacitor which quickly brings the potential down across the surface as shown in Fig. \ref{fig:3n}. An analogous structure also exists in heavy nuclei \cite{Pomeranchuk, Migdal_1977} with charge $Z\gg1/\alpha^{3/2}$.

To make the analysis more quantitative, one needs to know the specific potential profile in this region.
Near the cluster surface, we can approximately use a plane solution of $\varphi(r)$, i.e., ignore the $2/r$ term on the left side of Eqs. (\ref{eq:9a}) and (\ref{eq:9b}). This kind of solution for $r\gtrsim R$ is given by Eq. (\ref{eq:potential_nonlinear}) with $x=r-R\ll R$ which is the distance to the surface and the characteristic decay length $d\ll R$ is given by Eq. (\ref{eq:32}) which in the dimensional form is
\begin{equation}
d=\frac{C_{11}}{A^{1/4}}\left(\frac{b}{a}\right)^{1/4}\frac{a}{(n_0a^3)^{7/12}},
\label{eq:13n}
\end{equation}
where $C_{11}\approx 2$. By expressing $n_0$ in terms of $Z$ and $R$, we get $d/R\propto\left(Z^*/Z\right)^{7/12}\ll 1$ at $Z\gg Z^*$.

Correspondingly, the radial electron concentration at $x\gtrsim R$ is given by
\begin{equation}
\begin{aligned}
n(r)r^2&=\frac{C_{12}}{A^{3/7}}\frac{1}{b^3}\left(\frac{b}{a}\right)^{24/7}\left(\frac{a}{x+d}\right)^{12/7}r^2\\
&\approx\frac{C_{12}}{A^{3/7}}\frac{1}{b^3}\left(\frac{b}{a}\right)^{24/7}\left(\frac{a}{x+d}\right)^{12/7}R^2,
\end{aligned}
\label{eq:12n}
\end{equation}
where $r\approx R$, $C_{12}\approx1$.

Since the ``double-layer" structure resembles a plane capacitor, near the surface, the potential drops practically linearly with the radius. Using Eq. (\ref{eq:potential_nonlinear}), one can get $\varphi(r)\approx [1-8x/7d]\varphi(R)$ at $0<x=r-R\ll d$, which gives the electric field $8\varphi(R)/7d$ inside the ``double layer". At $r<R$, this electric field persists and gives $\varphi(r)\approx [1+8(-x)/7d]\varphi(R)$ at $0<-x=R-r\ll d$. As $r$ further decreases, the positive layer ends and the potential crosses over to the constant value $\varphi_0$  given by $n(\varphi_0)=n_0$ using Eq. (\ref{eq:5n}).

According to Eq. (\ref{eq:potential_nonlinear}), when $x=r-R$ is comparable to $R$ and the plane approximation is about to lose its validity, $\varphi(r)$ is $\propto(R/a)^{-8/7}$. It is weak enough to match the low electric potential $\varphi_{out}(r)\propto(R/a)^{-8/7}$ caused by the renormalized charge $Z_n\sim Z^*$ at $r\sim R$. The plane solution then crosses over to the potential $\varphi_{out}(r)\propto {Z^*}^3/r^5$ which is the asymptotic form at large distances.

\begin{figure}[h]
\includegraphics[width=0.5\textwidth]{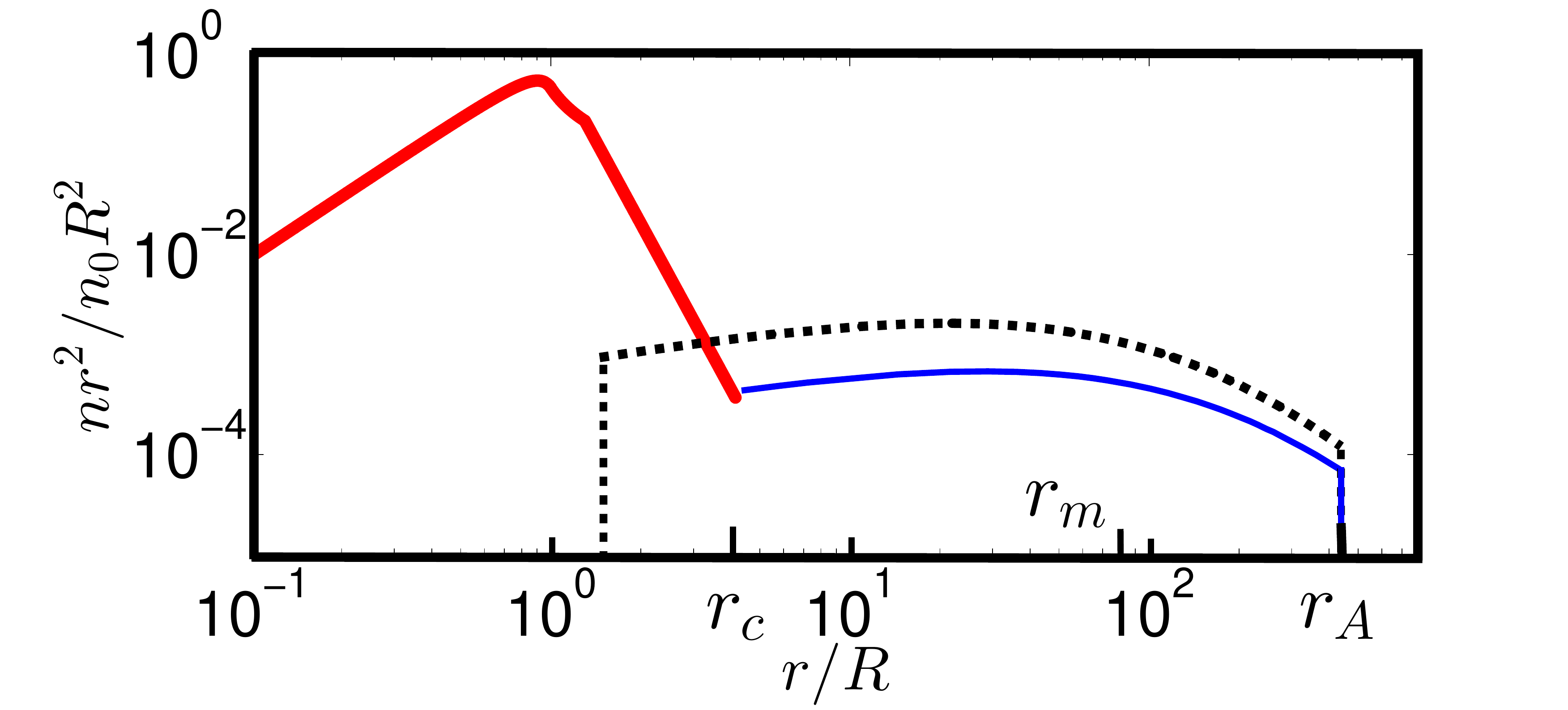}\\
\caption{(Color online) Radial electron concentration $n(r)r^2$ as a function of radius $r$ for the two-scale Thomas-Fermi atom formed by the super-critical cluster of donors (solid lines). The thick solid line (red) represents the inner collapsed electrons at $r<r_c$ where the dielectric response is nonlinear. The thin solid line (blue) shows the electrons belonging to the outer shell which form the standard Thomas-Fermi atom with the renormalized nucleus of charge $Z^*$ at $r>r_c$, where the dielectric response is linear. This electron gas ends at the Bohr radius $r_A=\kappa b$ while most of them are at the radius $r_m=\kappa b/{Z^*}^{1/3}$. For contrast, the dashed line (black) denotes the electrons forming a conventional Thomas-Fermi atom \cite{Thomas_Fermi,Landau} with a nucleus of charge $Z$ when $P_0$ is infinity and there's no range with nonlinear dielectric response. The reduction of electron density in the outer shell of electrons due to the collapse is substantial. The reason this is not immediately seen from the difference of height between the dashed line (black) and the thin solid line (blue) is that we use a logarithmic scale here. $n_0$ is defined as $3Z/4\pi R^3$. This graph is plotted at $b=0.35\,\AA$, $a=3.9\,\AA$, $A=0.9$, $R=4.4a$, $\kappa=20000$, $n_0=0.8/a^3$.}\label{fig:3a}
\end{figure}

A schematic plot of the potential energy $U(r)=-e\varphi(r)$ as a function of radius $r$ is shown in Fig. \ref{fig:3n} by the thin solid line (red). The corresponding radial distribution of electrons is shown in Fig. \ref{fig:3a} by the thick solid line (red).

$\phantom{}$

So far, we have got a $1/r^5$ potential $\varphi(r)$ and $1/r^{11/2}$ radial electron concentration $n(r)r^2$ at $r\gg R$ in both weak and strong charge renormalization cases. However, as the electron density decreases to certain extent so that the Fermi wavelength $\lambda$ is comparable to the radius $r$, the gas is no longer degenerate and the Thomas-Fermi approach fails. Since $\lambda\simeq n(r)^{-1/3}$, we get this radius $r^*\simeq Z^*a$ at $Z\gg Z^*$. One should then return to the Schrodinger equation used for a single electron. Since the uncertainty principle estimates that the kinetic energy decays as $1/r^2$ while the potential energy is $\propto -{Z^*}^3/r^5$, the potential energy is smaller than the kinetic energy in magnitude at $r>r^*$, which means electrons can not stay at radii larger than $r^*$. One can also find that using the Thomas-Fermi solution $\varphi_{out}(r)$, the total electron number calculated at $r>r^*$ is $\sim 1$, which again indicates there is no electron at $r>r^*$ considering the discreteness of electron charge. As a result, the $1/r^{11/2}$ tail of radial electron concentration will not continue to infinity but stop at radius $r^*$. This is a semi-classical result. Quantum mechanical analysis shows that the electron density does not go to zero right at $r^*$ but decays exponentially after this point. Since this decay is fast and brings very small corrections to the edge of the inner electron gas, we do not consider it here.

At $\kappa=\infty$, the rest of the electrons are at the infinity so that we are dealing with a positive ion with charge $Z^*$. At finite but very large $\kappa$, at certain distance from the cluster, the field is so small that $P>\sqrt{4\pi/\kappa A}P_0$ is no longer satisfied and the linear dielectric response is recovered. Things then become quite familiar. Electrons are mainly located between $r_1=\kappa b/Z^*$ and $r_A=\kappa b$ with the majority at radius $r_m=\kappa b/{Z^*}^{1/3}$ as given by the Thomas-Fermi model \cite{Landau}. Although quantum mechanics gives a nonzero electron density at $r<r_1$, the number of total electrons within this radius is only $\sim 1$ and can be ignored. So approximately, when $r_1\gg r^*$, i.e., $\kappa\gg (Z^*)^2\simeq(R/a)^{18/7}$, there's a spatial separation between inner collapsed electrons and outer ones that form the usual Thomas-Fermi atom with the renormalized nucleus. When $\kappa$ is not so big, such separation is absent, which actually happens more often in real situations. The inner tail then connects to the outer electrons with the Thomas-Fermi approach valid all the way and the dielectric response becomes linear at $r=r_c\propto a\kappa^{1/4}{Z^*}^{1/2}$. One should note, as long as $\kappa$ is large enough to satisfy $r_m\gg R$ which gives $\kappa\gg (R/a)^{10/7}$, the majority of the outer electrons located at $r_m$ do not intrude into the cluster or the highly screening double-layer structure near the cluster surface. The charge renormalization process remains undisturbed and the total net charge seen by outer electrons is still $Z^*$. The corresponding radial electron concentration $n(r)r^2$ is shown in Fig. \ref{fig:3a}. At $\kappa\ll(R/a)^{10/7}$, in most of the space the dielectric response is linear. In that case, almost all electrons reside in the cluster with only some spill-out near the surface. The positive and negative charges are uniformly distributed inside the cluster as described by the Thompson ``jelly" model.

\section{Cylindrical Donor Clusters}
\label{sec:cylindrical}
In some cases, the donor clusters are more like long cylinders than spheres. Then, a cluster is described by the linear charge density $\eta e$ while its radius is still denoted as $R$. We use a cylindrical coordinate system with the $z$ axis along the axis of the cylinder cluster and $r$ as the distance from the axis.
We show that when the charge density $\eta e$ is larger than certain value $\eta_c e$, electrons begin to collapse into the cluster and the charge density is weakly renormalized. When $\eta$ exceeds another value $\eta^*\gg\eta_c$, the renormalization becomes so strong that the net density $\eta_n$ remains $\simeq \eta^*$ regardless of the original density $\eta$.
Our problem is similar to that of the charged vacuum condensate near superconducting cosmic strings \cite{Vilenkin_1999}, and is also reminiscent of the Onsager-Manning condensation  in salty water \cite{Onsager_1967, *Manning_1969}\footnote{For example, in salty water, the negative linear charge density of DNA is renormalized from $\simeq-4e/l_B$ to the universal net value $-e/l_B$ due to the condensation of Na$^+$ ions onto the DNA surface. Here $l_B=e^2/\kappa_wk_B T\simeq7\,\AA$ where $\kappa_w=81$ is the dielectric constant of water and $T$ is the room temperature.}.

\emph{Renormalization of Linear Charge Density.}---
For a uniformly charged cylindrical cluster with a linear charge density $\eta e$, similar to what we did in Sec. \ref{sec:spherical}, we get $D(r)=2\eta(r)e/r$, where $\eta(r)$ is the total linear charge density enclosed in the cylinder of radius $r$ and $\eta(r)=\eta r^2/R^2$  at $r<R$ and $\eta(r)=\eta$ at $r>R$.  We then can calculate the electric field using Eqs. (\ref{eq:Gauss}) and (\ref{eq:electric_field_definition}) and get the electric potential $\varphi(r)$ as:
\begin{subequations}
\begin{align}
\varphi(r)&=\frac{A}{P_0^2}\left(\frac{\eta e}{2\pi}\right)^3\left(\frac{3}{4}\frac{1}{R^2}-\frac{1}{4}\frac{r^4}{R^6}\right), 0<r<R\\
\varphi(r)&=\frac{A}{P_0^2}\left(\frac{\eta e}{2\pi}\right)^3\frac{1}{2r^2}, \quad\quad\quad\quad\quad\quad\quad R<r\label{eq:14b}
\end{align}
\end{subequations}
with $\varphi(r=\infty)$ chosen to be $0$. The corresponding potential energy $U(r)=-e\varphi(r)$ is shown in Fig. \ref{fig:5n} by the thick solid line (blue). Using the Schrodinger equation and setting the momentum $p\simeq\hbar/2r$, we find that the tightly bound states of electrons, in which electrons are strongly confined within the cluster (at $r<R$), exist only when $\eta>\eta_c$ where
\begin{equation}
\eta_c\approx2\pi\left(\frac{b}{Aa}\right)^{1/3}\frac{1}{a},
\end{equation}
which, contrary to $Z_c$ obtained in the spherical case, does not depend on $R$. Electrons begin to collapse into the cluster at $\eta>\eta_c$ and in the beginning they are located near the axis where the potential energy is lowest (see Fig. \ref{fig:5n}). With increasing $\eta$, the electron density grows and one can adopt the Thomas-Fermi description. Using Eq. (\ref{eq:5n}) and (\ref{eq:14b}), one gets the electron density $n(r)\propto 1/r^3$ at $r>R$ and the total number of collapsed electron per unit length is
\begin{equation}
\theta=\int_0^\infty n(r)2\pi rdr=0.5\eta\left(\frac{\eta }{\eta^*}\right)^{7/2}\propto\eta^{9/2},
\label{eq:16n}
\end{equation}
where
\begin{equation}
\eta^*=\frac{1}{a}\left[2\pi\left(\frac{b}{Aa}\right)^{1/3}\right]^{9/7}\left(\frac{R}{a}\right)^{2/7}.
\label{eq:18n}
\end{equation}
%where the coefficient $g_2$ is of order 1.
The net charge density $\eta_ne$ is then renormalized to
\begin{equation}
\eta_n=\eta-\theta=\eta\left[1-0.5\left(\frac{\eta}{\eta^*}\right)^{7/2}\right].
\end{equation}

\begin{figure}[h]
\includegraphics[width=0.5\textwidth]{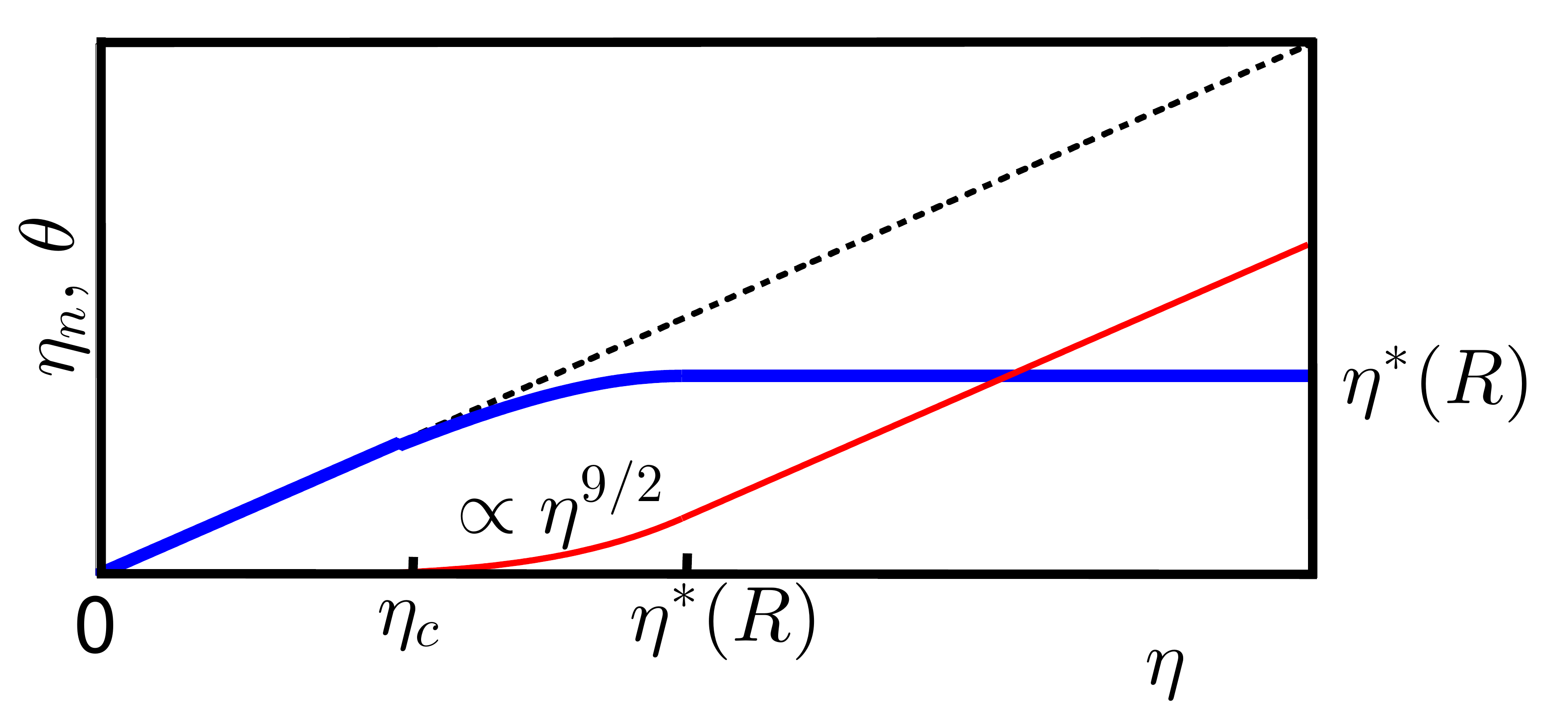}\\
\caption{(Color online) Number of collapsed electrons per unit length $\theta$ and renormalized net linear charge density $\eta_n$ as a function of the cluster linear charge density $\eta$. The thick solid line (blue) shows $\eta_n(\eta)$. The thin solid line (red) represents $\theta(\eta)$. The dashed line (black) is a guide-to-eye with $\eta_n=\eta$. $\theta(\eta)\propto \eta^{9/2}$ at $\eta_c\ll\eta\ll\eta^*$.}\label{fig:4n}
\end{figure}

\begin{figure}[h]
$\begin{array}{l}
\includegraphics[width=0.5\textwidth]{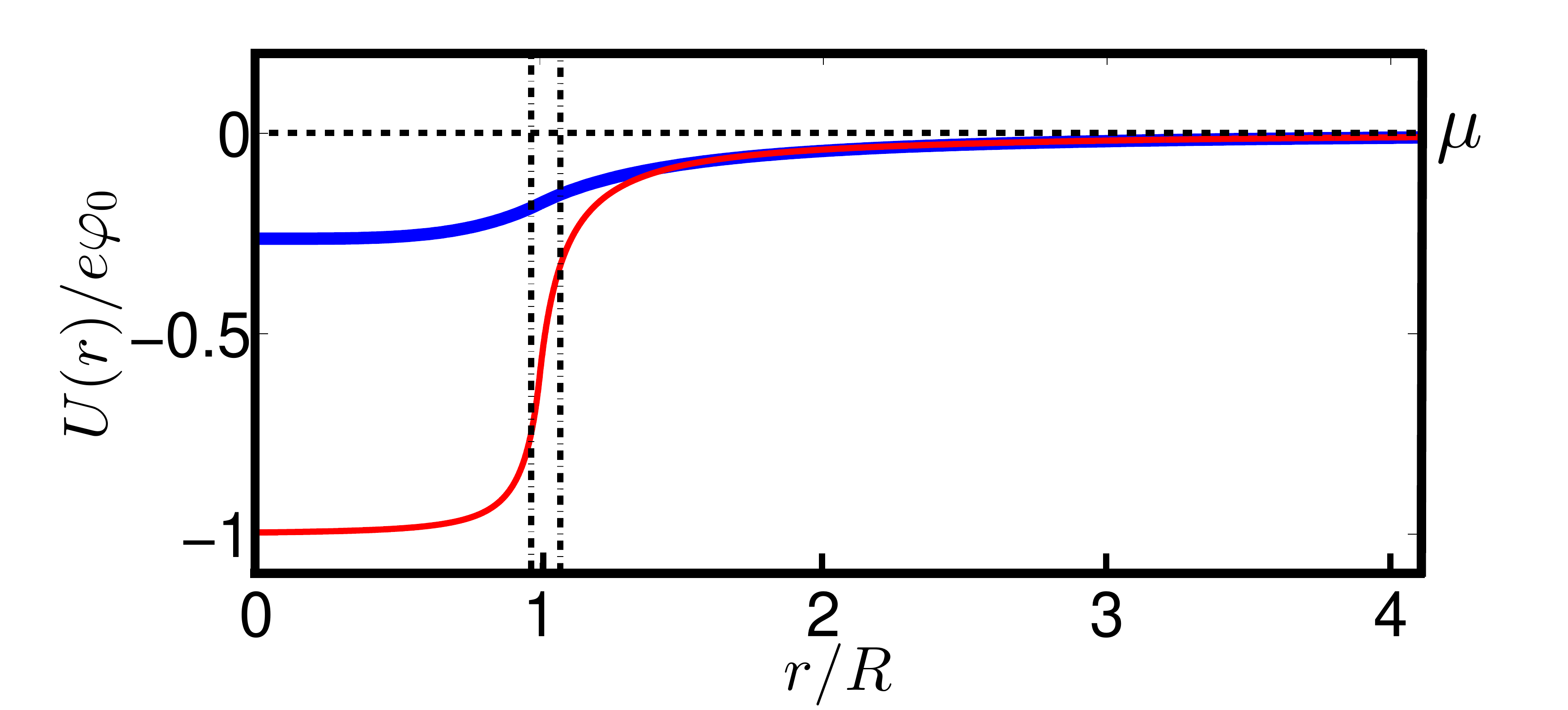}\\
\end{array}$
\caption{(Color online) Potential energy of electrons $U(r)=-e\varphi(r)$ caused by a cylindrical donor cluster of radius $R$ as a function of the distance $r$ from the cluster center. $\varphi_0$ is defined as $n(\varphi_0)=n_0$, where $n_0=\eta/\pi R^2$, $n(r)$ is a function of $\varphi(r)$ given by Eq. (\ref{eq:5n}). The thick solid line (blue) represents the potential profile of a cluster of charge density $\eta\lesssim \eta^*$ which is in the regime of weak renormalization of charge. The thin solid line (red) represents the potential of a cluster with $\eta\gg \eta^*$ which is in the strong renormalization regime. The two vertical dotted lines show edges of the ``double-layer" structure of width $\sim d \ll R$. The horizontal dashed line (black) indicates the position of the chemical potential $\mu= 0$.}
\label{fig:5n}
\end{figure}

At $\eta\ll \eta^*$, the renormalization of charge density is weak and $\eta_n$ grows with $\eta$. At $\eta>\eta^*$, the renormalization effect becomes strong. Most of the cluster is then neutralized by electrons and the final net density $\eta_n$ is much smaller than $\eta $. Following the logics similar to that in the spherical case, and by using Eq. (\ref{eq:16n}), one can show that $\eta_n$ reaches a saturation value of $\eta^*$ at $\eta\gg \eta^*$.% where $\gamma_2$ is a numerical coefficient of order 1.
$ $ The dependence of $\eta_n$ on $\eta$ is shown in Fig. \ref{fig:4n}.

\emph{Radial Distribution of Electrons.}---
At $\eta\gg \eta^*$, there are lots of collapsed electrons inside the cluster where $n(r)=n_0=\eta/\pi R^2$ and the potential energy is low. Again, there's a ``double-layer" structure on the surface that provides steep growth of potential energy with $r $ at $r=R$.
Close to the cylinder surface at $0<x=r-R\ll R$, as for the sphere, we can approximately use a plane solution of $\varphi(r)$ as given by Eq. (\ref{eq:potential_nonlinear}). The expression of the characteristic decay length $d$ is also the same as in Eq. (\ref{eq:13n}).
When $x=r-R$ is comparable to $R$, the plane solution crosses over to the fast decaying potential $\propto 1/r^2$ as given by Eq. (\ref{eq:14b}) with $\eta$ replaced by $\eta_n\simeq\eta^*$. %The corresponding radial electron density $n(r)r$ is expressed by Eq. (\ref{eq:19}) also with $\eta$ replaced by $\eta_n$.
A schematic plot of the potential energy $U(r)=-e\varphi(r)$ is shown in Fig. \ref{fig:5n}.% while the corresponding radial distribution of electrons $n(r)r$ is shown in Fig. (\ref{fig:6})

This potential produces a universal tail of electron density $n(r)\sim 1/r^3$. The corresponding radial electron concentration $n(r)r$ is $\sim 1/r^2$. Since the Fermi wavelength $\lambda\simeq n(r)^{-1/3}$, we get $\lambda\sim r$, i.e., the Thomas-Fermi approach is only marginally valid. The collapsed electrons extend until the linear dielectric response is recovered and then connect to the outer electrons.

\section{Finite-temperature effect in spherical donor clusters and its experimental implications}
\label{sec:temp}

So far, we dealt with zero temperature. At a finite temperature $T$, the neutral cluster atom can get ionized due to the entropy gain of ionized electrons.
The donor cluster atom becomes a positive ion with charge $Z_i(T)e$. Recall that the TF approach is valid at all distances until $r_A=\kappa b$. The first ionization energy of the cluster atom is then negligible ($=e^2/\kappa^2b\simeq 10^{-7}$ eV). So the cluster atoms are always partially ionized.
Our goal below is to find this ionization charge for the spherical donor clusters. Similar analyses for the cylindrical clusters can be found in Ref. \onlinecite{HRS} and are not repeated here.

We assume that we have a small but finite three-dimensional concentration $N$ of spherical clusters and the charge $Z_i(T) < Z^*$, i. e., the outer electron shell is still incompletely ionized. Such a cluster can bind electrons with an ionization energy $Z_i(T)^{2}e^2/\kappa^2 b$. We can find $Z_i(T)$ by equating this energy with the decrease in the free energy per electron $k_BT\ln(n_0/n)$ due to the entropy increase (the entropy increase can be derived according to $\S$ 104 of Ref. \onlinecite{Landau_stat}), where $k_B$ is the Boltzmann constant,
$n=Z_i(T)N$ is the concentration of ionized electrons and $n_0=2/\lambda^3$ with $\lambda=\sqrt{2\pi\hbar^2/m^*k_BT}$ as the DeBroglie wavelength of free electrons at temperature $T$. At $\kappa=20000$, $b=0.29\,\AA$, $m^*=1.8m_e$ where $m_e$ is the electron mass and $N=10^{15}$ cm$^{-3}$ (estimated from that the concentration of total donor electrons is around $10^{18}$ cm$^{-3}$ and each cluster contributes $\sim 300$ donor electrons), we get $Z_i(T)\gtrsim Z^*$ at $T\gtrsim 8$ K with $Z^*=100$ which is a reasonable estimate. This shows that the outer electrons are completely ionized at temperatures that are not too low. For the inner core electrons, the dielectric response is nonlinear and the attractive potential is stronger. So the ionization energy is higher $\simeq A(Z^*e/4\pi)^3/5P_0^2R^5$ for electrons at $r\simeq R$. At $R=4a$, it is found that only at $T>450$ K can
the inner electrons be ionized by a considerable quantity (the $1/r^{15/2}$ tail is completely stripped then). So the inner electrons are robust against the thermal ionization.

Experimentally, charged clusters can be created controllably on the surface of LAO/STO structure when the LAO layer is of subcritical thickness $\lesssim 3$ unit cell \cite{Cen_2008, Cen_2010}. A conducting atomic force microscope (AFM) tip is placed in contact with the top LaAlO$_3$ (LAO) surface and biased at certain voltage with respect to the interface, which is held at electric ground. When the voltage is positive, a locally metallic interface is produced between LAO and STO where some positive charges are accumulated in the shape of a disc. The same writing process can also create a periodic array of charged discs.

Let us first concentrate on a disc of positive charge created in this manner on the STO surface. Close to the surface and in the bulk STO, one should apply the plane solution given by Eq. (\ref{eq:conventration_nonlinear}). When the distance $r$ from the disc center is large, i.e., $r\gg R$, the disc behaves like a charged sphere. Our results for a sphere are still qualitatively correct in this case.

In a periodic array of highly charged discs with period $2L$ (see Fig. 3a in Ref. \onlinecite{Cen_2008}), the linear concentration of free electrons responsible for the conductance at a very low temperature is of the order of $n(L)L^2$, where $n(r)$ is the electron density around a spherical donor cluster given by Sec. \ref{sec:spherical}. When the overlapping parts between neighboring discs belong to the outer electron shells, the corresponding density at $r=L$ is that of a Thomas-Fermi atom with charge $Z^*$. In this situation, the overlapping external atmosphere forms conductive ``bridges" between discs at low temperature. When $T$ increases, however, the outer electrons are ionized and the bridges are gone. These free electrons spread out over the bulk STO. At $T\lesssim 30$ K, electrons are scattered mainly by the Coulomb potential of donors and the corresponding mobility decreases with a decreased electron velocity. For the electrons ionized into the vast region of the bulk STO, they are no longer degenerate, so their velocity becomes much smaller at relatively low temperature. This results in a much smaller mobility of the ionized electrons than those bound along the chain. Their contribution to the conductivity is thus negligible. The system becomes more resistive due to the ionization and one can observe a sharp decrease of the conductivity along the chain.

\section{Conclusion}
\label{sec:conclusion}
In this paper we have studied the potential and electron density depth profiles in surface accumulation layers in crystals with a large and nonlinear dielectric response such as SrTiO$_3$ (STO) in the cases of planar, spherical and cylindrical geometries. We use the Landau-Ginzburg free energy expansion for the dielectric response of STO and adopt the Thomas-Fermi approximation for its electron gas.

For the planar geometry we predict an electron density profile $n(x) \propto (x+d)^{-12/7}$, where $d \propto D_0^{-7/5} $ with $D_0$ as the surface induction and $x$ as the distance from the interface. Here we skipped comparison of this result with the experimental data which can be found in Ref. \onlinecite{RS}.  The data generally show a reasonable agreement with our predictions. We extend our studies from a single accumulation layer to overlapping ones, and also investigate the electron ``spill-out" from a heavily $n$-type doped STO (NSTO) into a moderately $n$-type doped STO.

In the second part of this paper we study the collapse of electrons onto spherical and cylindrical donor clusters. Such `` fall-to-the-center" originates from the very fast decrease of the electron potential energy near the cluster which is $\propto (-1/r^5)$ in the spherical case and in turn is a result of the strongly nonlinear dielectric response of STO. This leads to a very unusual two-scale shape of the electron density around the cluster. We show how one can verify such a shape experimentally.

$\phantom{}$
\vspace*{2ex} \par \noindent
{\em Acknowledgments.}

We are grateful to B. Jalan, E. B. Kolomeisky, M. Schecter, and S. Stemmer for helpful discussions. This work was supported primarily by the National Science Foundation through the University of Minnesota MRSEC under Award No. DMR-1420013.

%\bibliography{/home/reich/Science/Papers/papers.bib,papers2.bib}
%

\end{document}